%% file: ms.tex
\RequirePackage{fix-cm}
\documentclass[smallextended,a4paper]{svjour3}       

\smartqed

\usepackage{graphicx}
\usepackage{xcolor}
\usepackage{natbib}
\usepackage{amsmath}
\usepackage{amssymb}
\usepackage{bm}

\newcommand{\ie}{i.e.,\ }
\newcommand{\eg}{e.g.,\ }

\newcommand{\bra}[1]{\langle #1\rangle}
\def \Rey  {\mbox{Re}}

\def \Ca  {\mbox{Ca}}
\def \muf {\mu_0}

\def \vf {u^{f1}}

\def \phis {\phi}

\newcommand{\secref}[1]{\mbox{section \ref{#1}}}

\newcommand{\equref}[1]{\mbox{equation (\ref{#1})}}

\newcommand{\tabref}[1]{\mbox{table \ref{#1}}}

\newcommand{\figrefC}[2][]{\mbox{Figure \ref{#2}(#1)}}
\newcommand{\figref}[2][]{\mbox{figure \ref{#2}(#1)}}
\newcommand{\figrefSC}[1]{\mbox{Figure \ref{#1}}}
\newcommand{\figrefS}[1]{\mbox{figure \ref{#1}}}

\definecolor{cinnamon}{rgb}{0.82, 0.41, 0.12}

\journalname{Acta Mechanica}

\AtBeginDocument{%
  \setlength{\oddsidemargin}{\dimexpr(\paperwidth-\textwidth)/2-1in}%
  \setlength{\evensidemargin}{\oddsidemargin}%
  \setlength{\topmargin}{%
    \dimexpr(\paperheight-\textheight)/2-\headheight-\headsep-1in}%
}

\begin{document}

\title{Numerical simulations of emulsions in shear flows}

\author{Marco E. Rosti \and Francesco De Vita \and Luca Brandt}
\institute{Linn\'{e} Flow Centre and SeRC, KTH Mechanics, Stockholm, Sweden \\ Email address for correspondence: merosti@mech.kth.se}

\date{Received: date / Accepted: date}

\maketitle

\begin{abstract}
We present a modification of a recently developed volume of fluid method for multiphase problems \citep{ii_sugiyama_takeuchi_takagi_matsumoto_xiao_2012a}, so that it can be used in conjunction with a fractional step-method and fast Poisson solver, and validate it with standard benchmark problems. We then consider emulsions of two-fluid systems and study their rheology in a plane Couette flow in the limit of vanishing inertia. We examine the dependency of the effective viscosity $\mu$ on the volume-fraction $\Phi$ (from $10\%$ to $30\%$) and the Capillary number $Ca$ (from $0.1$ to $0.4$) for the case of density and viscosity ratio $1$. We show that the effective viscosity decreases with the deformation and the applied shear (shear-thinning) while exhibits a non-monotonic behavior with respect to the volume fraction. We report the appearance of a maximum in the effective viscosity curve and compare the results with those of suspensions of rigid and deformable particles and capsules. We show that the flow in the solvent is mostly a shear flow, while it is mostly rotational in the suspended phase; moreover this behavior tends to reverse as the volume fraction increases. Finally, we evaluate the contributions to the total shear stress of the viscous stresses in the two fluids and of the interfacial force between them.
\end{abstract}

\section{Introduction} \label{sec:introduction} 
In the last decades, developments in colloidal science have proven to be crucial for fabrication of functional materials. Particles or droplets, with typical scales of micron, are manipulated to self-organize into controlled patterns with high precision, which then form basic building blocks for more complex structures \citep{sacanna_pine_2011a}. One important aspect to consider during the synthesis and assembly of innovative materials is the rheological behavior of the system \citep{mewis_wagner_2012a}. Indeed, the material properties will depend on the distribution of the dispersed phase and thus a more accurate control on the production process can help to generate  materials of desired properties \citep{xia_gates_yin_lu_2000a}.

A great amount of work has been done in the past to study rigid and deformable particle suspensions \citep{freund_2014a, takeishi_imai_ishida_omori_kamm_ishikawa_2016a, alizad-banaei_loiseau_lashgari_brandt_2017a, rosti_brandt_mitra_2018a, rosti_brandt_2018a}. In his pioneering work, \citet{einstein_1956a} showed  that, in the limit of vanishing inertia and for dilute rigid particle suspensions, the viscosity is a linear function of the particle volume fraction. Although \citet{batchelor_green_1972a} and \citet{batchelor_1977a} added a second order correction, all existing analytical relations are not valid for moderately high concentrations and one needs to resort to empirical fits. One of the available empirical relations that provides a good description of the rheology at zero Reynolds number both in the high and low concentration limits is the Eilers fit \citep{ferrini_ercolani_de-cindio_nicodemo_nicolais_ranaudo_1979a, zarraga_hill_leighton-jr_2000a, singh_nott_2003a, kulkarni_morris_2008a}. Recently, inertia and deformation have been shown to introduce deviations from the behavior predicted by the different empirical fits, and these effect can be related to an increase and decrease of an effective volume fraction \citep{picano_breugem_mitra_brandt_2013a, rosti_brandt_mitra_2018a}.

Less attention has been given to emulsions, which are instead the object of the present work. Emulsions are biphasic liquid-liquid systems in which the phases are separated by a deformable interface subject to interfacial surface tension. These can be found in a variety of applications, ranging from advanced materials processing, waste treatment, enhanced oil recovery, food processing, and pharmaceutical manufacturing. Similarly to particle suspensions, it is often desirable to predict or manipulate the rheology and microstructure of emulsions, which in general exhibit highly varied rheological behaviors \citep{mason_1999a}; however, there has been limited progress towards the creation of theoretical models that can reliably predict the rheology and microstructure of such flows \citep{loewenberg_hinch_1996a, loewenberg_1998a} and for years measurements of emulsion rheology were not quantitatively understood because of the complexity of the phenomenology involved and the difficulty to properly choose materials with controlled properties \citep{mason_1999a}. Thus, numerical simulations can help to fill this gap, and indeed there has been considerable progress in the development of numerical simulations able to study such multiphase flows \citep{prosperetti_tryggvason_2009a, tryggvason_scardovelli_zaleski_2011a}.

Different techniques have been proposed to numerically tackle the problem at hand. The so-called front-tracking method is an Eulerian/Lagrangian method, used to simulate viscous, incompressible, immiscible two-fluid systems, first developed by \citet{unverdi_tryggvason_1992a} and \citet{tryggvason_bunner_esmaeeli_juric_al-rawahi_tauber_han_nas_jan_2001a}. When dealing with moving and deformable boundaries, an alternative approach are the so-called front-capturing methods, which are fully Eulerian and handle topology changes automatically. A strong advantage of these methods is that they are easier to parallelize than their Lagrangian counterpart. Eulerian interface representations include essentially the volume of fluid (VOF) \citep{scardovelli_zaleski_1999a} and level-set (LS) \citep{sussman_smereka_osher_1994a, sethian_1999a, sethian_smereka_2003a} methods. The volume of fluid method defines different fluids with a discontinuous color function, and its main advantage is the intrinsic mass conservation; however, it suffers from an inaccurate computation of the interface properties, such as normals and curvatures \citep{francois_cummins_dendy_kothe_sicilian_williams_2006a, cummins_francois_kothe_2005a}. Differently from the volume of fluid, the level-set method prescribes the interface through a continuous function which usually takes the form of the signed distance to the interface. Thus, normals and curvatures can be readily and accurately computed, while mass loss or gain may occur. In this work we will employ the volume of fluid method.

In a conventional VOF method, the interface separating different fluids is piece-wisely reconstructed in each numerical cell by straight line segments, which are then used to calculate the numerical fluxes necessary to update the local volume of fluid function. This geometric reconstruction effectively eliminates the numerical diffusion that smears out the compactness of the transition layer of the interface. Different methodologies have been proposed to accurately recover the exact surface geometry from the discretized VOF function: the simple line interface calculation (SLIC) method \citep{noh_woodward_1976a}, the piecewise linear interface calculation (PLIC) \citep{youngs_1982a, youngs_1984a}, the latter being further modified by several authors \citep{puckett_almgren_bell_marcus_rider_1997a, rider_kothe_1998a, harvie_fletcher_2000a, aulisa_manservisi_scardovelli_zaleski_2003a, pilliod-jr_puckett_2004a}. Another technique is the tangent of hyperbola for interface capturing (THINC) method \citep{xiao_honma_kono_2005a}, which avoids the explicit geometric reconstruction by using a continuous sigmoid function rather than the Heaviside function, thus allowing a completely algebraic description of the interface; this enables the computation of the numerical flux partially analytically. An improvement was proposed by combining the original THINC method with the first-order upwind scheme in the so-called THINC/WLIC (THINC/weighted linear interface capturing) method \citep{yokoi_2007a}. Recently, the method has been further developed in the multi-dimensional THINC (MTHINC) method where the fully multi-dimensional hyperbolic tangent function is used to reconstruct the interface \citep{ii_sugiyama_takeuchi_takagi_matsumoto_xiao_2012a}. The numerical fluxes can be directly evaluated by integrating the hyperbolic tangent function and the normal vector, curvature and approximate delta function can be directly obtained from the derivatives of the function. Moreover, the scheme does not require the geometric reconstruction and a curved (quadratic) surface can be easily constructed as well.

\subsection{Outline}
In this work, we first present our numerical solver for multiphase incompressible flows and then employ it to study liquid-liquid systems (emulsions) in a plane Couette flow at low Reynolds number. The two fluids are Newtonian and satisfy the full incompressible Navier-Stokes equations. We compare our results with those of suspensions of rigid and deformable particles and with capsules, consisting of a second fluid enclosed by a thin elastic membrane. In \secref{sec:formulation}, we first discuss the flow configuration and governing equations, and then present the numerical methodology used and its validation. The rheological study of the emulsions is presented in \secref{sec:result}, where we also discuss the role of the different non-dimensional parameters governing the flow. Finally, a summary of the main findings and some conclusions are drawn in \secref{sec:conclusion}.

\section{Formulation} \label{sec:formulation}
\begin{figure}
  \centering
  \includegraphics[width=0.4\textwidth]{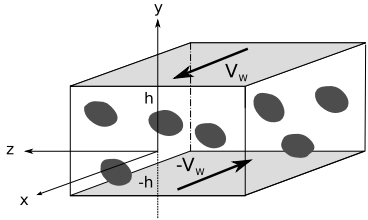}
  \caption{Sketch of the channel geometry and coordinate system adopted in this study.}
  \label{fig:sketch}
\end{figure}
We consider the flow of two incompressible viscous fluids, separated by an interface, in a channel with moving walls, \ie in a plane Couette geometry. \figrefSC{fig:sketch} shows a sketch of the geometry and the Cartesian coordinate system, where $x$, $y$ and $z$ ($x_1$, $x_2$, and $x_3$) denote the streamwise, wall-normal and spanwise coordinates, while $u$, $v$ and $w$ ($u_1$, $u_2$, and $u_3$) denote the corresponding components of the velocity vector field. The lower and upper impermeable moving walls are located at $y=-h$ and $y=h$, respectively, and move in opposite direction with constant streamwise velocity $\pm V_w$.

The two fluid motion is governed by the conservation of momentum and the incompressibility constraint, and the kinematic and dynamic interactions between the two fluid phases are determined by enforcing the continuity of the velocity and traction force at the interface between the two phases, \ie
\begin{equation}
\label{eq:bc}
u_i^{f1} = u_i^{f2} \;\;\;\;\;\ \textrm{and} \;\;\;\;\;\ \sigma_{ij}^{f1} n_j = \sigma_{ij}^{f2} n_j + \sigma \kappa n_i
\end{equation}
where the suffixes $^{f1}$ and $^{f2}$ are used to indicate the two phases, $\sigma_{ij}$ denotes the Cauchy stress tensor, $n_i$ the normal vector at the interface, $\kappa$ the interface curvature and $\sigma$ the surface tension (assumed here to be constant).

To numerically solve the two-phase interaction problem at hand, we use the volume of fluid method following \cite{ii_sugiyama_takeuchi_takagi_matsumoto_xiao_2012a}. We introduce an indicator (or color) function $H$ to identify each fluid phase so that $H = 1$ in the region occupied by the fluid $f1$ and $H = 0$ otherwise.
Considering that the fluid is transported by the flow velocity, we update $H$ in the Eulerian framework by the following advection equation written in divergence form:
\begin{equation} \label{eq:vofadvdiv}
  \frac{\partial \phi}{\partial t} + \frac{\partial u_i H}{\partial x_i} = \phi \frac{\partial u_i}{\partial xi},
\end{equation}
where $u_i$ is the local fluid velocity and $\phi$ the cell-averaged value of the indicator function.

Once $\phi$ is known, the two-fluid equations can be rewritten in the so called one-continuum formulation \citep{tryggvason_sussman_hussaini_2007a}, so that only one set of equations is solved over the whole domain. This is achieved by introducing a monolithic velocity vector field $u_i$, defined everywhere and found by applying the volume averaging procedure \citep{takeuchi_yuki_ueyama_kajishima_2010a, quintard_whitaker_1994b}. Thus, $u_i$ is governed by the following set of equations
\begin{subequations}
\label{eq:NS}
\begin{align}
\frac{\partial u_i}{\partial t} + \frac{\partial u_i u_j}{\partial x_j} &= \frac{1}{\rho} \left( \frac{\partial \sigma_{ij}}{\partial x_j} + f_i \right), \\
\frac{\partial u_i}{\partial x_i} &= 0,
\end{align}
\end{subequations}
where $\rho$ is the density, $f_i$ the surface tension force defined as $f_i=\sigma \kappa n_i \delta$, being $\delta$ the delta function at the interface, and $\sigma_{ij}$ is the stress written in a mixture form, \ie
\begin{equation}
\label{eq:phi-stress}
\sigma_{ij} = \left( 1 - \phi \right) \sigma_{ij}^{f1} + \phi \sigma_{ij}^{f2}.
\end{equation}
Note that, we have chosen $\phi$ to be the volume fraction of fluid $2$, \ie this is zero in the fluid $1$, whereas $\phi=1$ in the fluid $2$, and $0 \le \phi \le 1$ close to the interface. Both fluids are assumed to be Newtonian so that their stress tensors can be written as $\sigma_{ij} = -p \delta_{ij} + 2 \mu D_{ij}$, where $p$ is the pressure, $\delta_{ij}$ the Kronecker delta, $\mu$ the dynamic viscosity and $D_{ij}$ the strain rate tensor (defined as $D_{ij}=( \partial u_i/\partial x_j + \partial u_j/\partial x_i )/2$). Finally, the mixture density $\rho$ and dynamic viscosity $\mu$ are simply averaged in terms of the local $\phi$:
\begin{equation}  \label{eq:vofrhomu} 
\rho = \left( 1 - \phi \right) \rho^{f1} + \phi \rho^{f2} \;\;\;\;\;\ \textrm{and} \;\;\;\;\;\ \mu = \left( 1 - \phi \right) \mu^{f1} + \phi \mu^{f2}.
\end{equation}
Note that, in order to solve \equref{eq:NS} and \equref{eq:vofadvdiv}, we need to determine the indicator function $H$, the normal vector $n_i$ and the curvature $\kappa$.

\subsection{The MTHINC method}
The indicator function $H$ can be reconstructed in various ways; here, we use the multidimensional tangent of hyperbola for interface capturing (MTHINC) method, developed by \citet{ii_sugiyama_takeuchi_takagi_matsumoto_xiao_2012a}, where a multidimensional hyperbolic tangent function is used as an approximated indicator function. In particular, the indicator function $H$ is approximated as
\begin{equation} \label{eq:vofInd}
H \left( X, Y, Z \right) \approx \widehat{H} \left( X, Y, Z \right) = \frac{1}{2} \bigg( 1+ \tanh \big( \beta \left( P \left( X, Y, Z \right) + d \right) \big) \bigg),
\end{equation}
where $X, Y, Z \in \left[0, 1 \right]$ is a centered local coordinate system defined in each cell, $P$ is a three dimensional surface function, $\beta$ a sharpness parameter and $d$ a normalization parameter. The function $P$ can be either a linear function (a plane)
\begin{equation}
P \left( X, Y, Z \right) = a_{100} X + a_{010} Y + a_{001} Z,
\end{equation}
or a quadratic function (a curved surface)
\begin{equation}
\begin{split}
P \left( X, Y, Z \right) =& a_{200} X^2 + a_{020} Y^2 + a_{002} Z^2 + \\
& a_{110} XY + a_{011} YZ + a_{101} XZ + a_{100} X + a_{010} Y + a_{001} Z.
\end{split}
\end{equation}
The coefficients $a_{l,m,n}$ are determined algebraically by imposing the correct value of the three normal components $n_i$ and the six components of the Cartesian curvature tensor $l_{ij}=\left( \partial n_i/\partial x_j + \partial n_j/\partial x_i \right)/2$ for the function $P$ in each cell. Finally, the parameter $d$ is found by enforcing the following constraint:
\begin{equation} \label{eq:vofint}
\int_0^1 \int_0^1 \int_0^1 \widehat{H}~dX~dY~dZ = \phi.
\end{equation}
The integration can be performed analytically in one direction, and numerically in the other two directions by the two-point Gaussian quadrature.

The unit normal vector is defined as $n_i = m_i / \vert \nabla \phi \vert$, being $m_i$ the gradient of the volume of fluid function, \ie $m_i = \partial \phi/ \partial x_i$. Here, we compute $m_i$ using the usual Youngs approach \citep{youngs_1982a, youngs_1984a}, where first the values of the derivative at the cell corners are calculated, and then averaged to find the cell-center value. Once the normal vector is known, the curvature $\kappa$ can be easily found by taking the divergence of the normal vector, \ie $\kappa = - \partial n_i/\partial x_i$, and the surface tension force $f_i$ can be computed by the continuum surface force (CSF) model \citep{brackbill_kothe_zemach_1992a}, where the $1D$ approximate delta function $\delta$ is directly approximated by $\delta \approx \vert \nabla \phi \vert$. Thus, we obtain
\begin{equation}  \label{eq:voff}
f_i = \sigma \kappa n_i \delta \approx \sigma \kappa \frac{\partial \phi}{\partial x_i}.
\end{equation}

\subsection{Numerical discretisation}
The equation of motion are solved with an extensively validated in-house code \citep{picano_breugem_brandt_2015a, rosti_brandt_2017a, rosti_brandt_mitra_2018a, rosti_brandt_2018a}. The equations are solved on a staggered uniform grid with velocities located on the cell faces and all the other variables (pressure, stress and volume of fluid) at the cell centers. All the spatial derivatives are approximated with second-order centered finite differences, while the time integration is discussed hereafter.

First, the volume of fluid function is updated in time from the time-step $(n)$ to $(n+1)$ by solving \equref{eq:vofadvdiv}, following the procedure proposed by \citet{ii_sugiyama_takeuchi_takagi_matsumoto_xiao_2012a}. In particular, the time evolution of $\phi$ is calculated by evaluating the numerical fluxes sequentially in each direction, a robust and easy approach called directional splitting. Thus, \equref{eq:vofadvdiv} is discretised sequentially in the three Cartesian direction, \ie
\begin{equation} \label{eq:voftime1}
\begin{split}
\phi_{(*)}^{ijk}&=\phi_{(n)} -\frac{1}{\Delta x} \left( f^{i+\frac{1}{2},j,k}_{(n)} - f^{i-\frac{1}{2},j,k}_{(n)} \right) + \frac{\Delta t}{\Delta x} \phi^{i,j,k}_{(*)} \left( u_1^{i+\frac{1}{2},j,k}-u_1^{i-\frac{1}{2},j,k} \right), \\
\phi_{(**)}^{ijk}&=\phi_{(*)} -\frac{1}{\Delta x} \left( g^{i,j+\frac{1}{2},k}_{(*)} - g^{i,j-\frac{1}{2},k}_{(*)} \right) + \frac{\Delta t}{\Delta x} \phi^{i,j,k}_{(**)} \left( u_2^{i,j+\frac{1}{2},k}-u_2^{i,j-\frac{1}{2},k} \right), \\
\phi_{(***)}^{ijk}&=\phi_{(**)} -\frac{1}{\Delta x} \left( h^{i,j,k+\frac{1}{2}}_{(**)} - h^{i,j,k-\frac{1}{2}}_{(**)} \right) + \frac{\Delta t}{\Delta x} \phi^{i,j,k}_{(***)} \left( u_3^{i,j,k+\frac{1}{2}}-u_3^{i,j,k-\frac{1}{2}} \right),
\end{split}
\end{equation}
where the subscript in parenthesis indicates the time iteration, with $(n)$ and $(n+1)$ the old and new time steps, and $(*)$, $(**)$ and $(***)$ sub-iterations. Also, $\Delta t = t_{(n+1)} - t_{(n)}$ is the time-step, and $f$, $g$ and $h$ are the numerical fluxes defined later on. Note that, \equref{eq:voftime1} is implicit in the function $\phi$ in each sub-step. Next, we solve an additional equation in order to ensure the divergence-free condition of the fully multi-dimensional operator \citep{puckett_almgren_bell_marcus_rider_1997a, aulisa_manservisi_scardovelli_zaleski_2003a}:
\begin{equation} \label{eq:voftime2}
\begin{split}
\phi_{(n+1)}^{ijk}=\phi_{(***)} -\Delta t \bigg( &\phi_{(*)}^{ijk} \frac{u_1^{i+\frac{1}{2},j,k}-u_1^{i-\frac{1}{2},j,k}}{\Delta x} +  \\
&\phi_{(**)}^{ijk} \frac{u_2^{i,j+\frac{1}{2},k}-u_2^{i,j-\frac{1}{2},k}}{\Delta y} +\\
&\phi_{(***)}^{ijk} \frac{u_3^{i,j,k+\frac{1}{2}}-u_3^{i,j,k-\frac{1}{2}}}{\Delta z} \bigg).
\end{split}
\end{equation}
Finally, we need to specify how the numerical fluxes are treated. These are defined as the space/time integration of the product of the velocity $u_i$ and the indicator function $H$, which is substituted by its approximate counterpart $\widehat{H}$, \ie
\begin{equation} \label{eq:flux1}
\begin{split}
f_{(n)}^{i\pm\frac{1}{2},j,k} = \frac{1}{\Delta y \Delta z} \int_{\delta t_{(n)}} \int_{\Delta y} \int_{\Delta z} \left( u_1 \widehat{H} \right)^{i\pm\frac{1}{2},j,k} dy~dz~dt, \\
g_{(*)}^{i,j\pm\frac{1}{2},k} = \frac{1}{\Delta x \Delta z} \int_{\delta t_{(*)}} \int_{\Delta x} \int_{\Delta z} \left( u_2 \widehat{H} \right)^{i,j\pm\frac{1}{2},k} dx~dz~dt, \\
h_{(**)}^{i,j,k\pm\frac{1}{2}} = \frac{1}{\Delta x \Delta y} \int_{\delta t_{(**)}} \int_{\Delta x} \int_{\Delta y} \left( u_3 \widehat{H} \right)^{i,j,k\pm\frac{1}{2}} dx~dy~dt.
\end{split}
\end{equation}
The temporal integration can be replaced by a spatial integration along the upwind path on the velocity field. For example, in the $x$-direction the upstream path is $\Delta x_{+} = \left[ x^{i+\frac{1}{2}} - \Delta t u^{i+\frac{1}{2},j,k}, x^{i+\frac{1}{2}} \right]$ for $u_1^{i+\frac{1}{2}}\geq 0$ or $\Delta x_{-} = \left[ x^{i+\frac{1}{2}}, x^{i+\frac{1}{2}} - \Delta t u ^{i+\frac{1}{2},j,k} \right]$ for $u_1^{i+\frac{1}{2}}< 0$. Similarly in the other two directions. Thus, the numerical fluxes can be computed as
\begin{equation} \label{eq:flux2}
\begin{split}
f_{(n)}^{i+\frac{1}{2},j,k}=& \left\{ \begin{array}{cc}
\frac{1}{\Delta y \Delta z} \int_{\Delta x_{+}} \int_{\Delta y} \int_{\Delta z} \widehat{H}_{(n)}^{i,j,k} dx~dy~dz & \;\;\;\textrm{for}\;\;\; u_1^{i+\frac{1}{2},j,k}\geq 0 \\
-\frac{1}{\Delta y \Delta z} \int_{\Delta x_{-}} \int_{\Delta y} \int_{\Delta z} \widehat{H}_{(n)}^{i+1,j,k} dx~dy~dz & \;\;\;\textrm{for}\;\;\; u_1^{i+\frac{1}{2},j,k}< 0
\end{array} \right. \\
g_{(*)}^{i,j+\frac{1}{2},k}=& \left\{ \begin{array}{cc}
\frac{1}{\Delta x \Delta z} \int_{\Delta x} \int_{\Delta y_{+}} \int_{\Delta z} \widehat{H}_{(*)}^{i,j,k} dx~dy~dz & \;\;\;\textrm{for}\;\;\; u_2^{i,j+\frac{1}{2},k}\geq 0 \\
-\frac{1}{\Delta x \Delta z} \int_{\Delta x} \int_{\Delta y_{-}} \int_{\Delta z} \widehat{H}_{(*)}^{i,j+1,k} dx~dy~dz & \;\;\;\textrm{for}\;\;\; u_2^{i,j+\frac{1}{2},k}< 0
\end{array} \right. \\
h_{(**)}^{i,j,k+\frac{1}{2}}=& \left\{ \begin{array}{cc}
\frac{1}{\Delta x \Delta y} \int_{\Delta x} \int_{\Delta y} \int_{\Delta z_{+}} \widehat{H}_{(**)}^{i,j,k} dx~dy~dz & \;\;\;\textrm{for}\;\;\; u_3^{i,j,k+\frac{1}{2}}\geq 0 \\
-\frac{1}{\Delta x \Delta y} \int_{\Delta x} \int_{\Delta y} \int_{\Delta z_{-}} \widehat{H}_{(**)}^{i,j,k+1} dx~dy~dz & \;\;\;\textrm{for}\;\;\; u_3^{i,j,k+\frac{1}{2}}< 0
\end{array} \right.
\end{split}
\end{equation}
Similarly to \equref{eq:vofint}, the numerical integration can be performed analytically in one direction, and numerically in the other two by the two-point Gaussian quadrature.

Once the volume of fluid function has been updated, \ie $\phi^{n+1}$ is available, differently from what done by \citet{ii_sugiyama_takeuchi_takagi_matsumoto_xiao_2012a}, the time integration of \equref{eq:NS} is here performed with a fractional-step method \citep{kim_moin_1985a} where the evolution equation is advanced in time with a second-order Adam-Bashforth scheme and a Fast Poisson Solver is used to enforce zero divergence of the velocity field. Due to the non-uniformity of the density, the Poisson equation used to enforce a divergence-free velocity field results in an equation with variable coefficients, \ie
\begin{equation}
\frac{\partial}{\partial x_i} \left(\frac{1}{\rho} \frac{\partial p}{\partial x_i} \right) = \frac{1}{\Delta t}\frac{\partial \widehat{u}_i}{\partial x_i},
\end{equation}
where the pressure $p$ and density $\rho$ are evaluated at $n+1$ and the velocity $\widehat{u}_i$ is the non-divergence free predicted velocity. In order to employ an efficient FFT-based pressure solver with constant coefficients \citep[see also][]{dodd_ferrante_2014a, dodd_ferrante_2016a}, we use the following splitting of the pressure term \citep{dong_shen_2012a}:
\begin{equation}
\frac{1}{\rho} \frac{\partial p}{\partial x_i} \to \frac{1}{\rho_0} \frac{\partial p}{\partial x_i} +\left( \frac{1}{\rho} - \frac{1}{\rho_0} \right)  \frac{\partial \widetilde{p}}{\partial x_i},
\end{equation}
where $\rho_0$ is a constant density equal to the lowest density of the two phases, and $\widetilde{p}$ is an approximated pressure obtained by linear extrapolation, \eg  $\widetilde{p} = 2 p^{n}-p^{n-1}$. With this splitting, the Poisson equation can be rewritten as
\begin{equation} \label{eq:lsDodd}
\frac{\partial^2 p}{\partial x_i \partial x_i} = \frac{\rho_0}{\Delta t}\frac{\partial \widehat{u}_i}{\partial x_i} + \frac{\partial}{\partial x_i} \left[ \left( 1 - \frac{\rho_0}{\rho} \right) \frac{\partial \widetilde{p}}{\partial x_i}\right].
\end{equation}
Note that, also the correction step of the fractional-step method needs to be modified accordingly.

\subsection{Code validation} \label{sec:validation}
\begin{figure}
  \centering
  \includegraphics[width=0.24\textwidth]{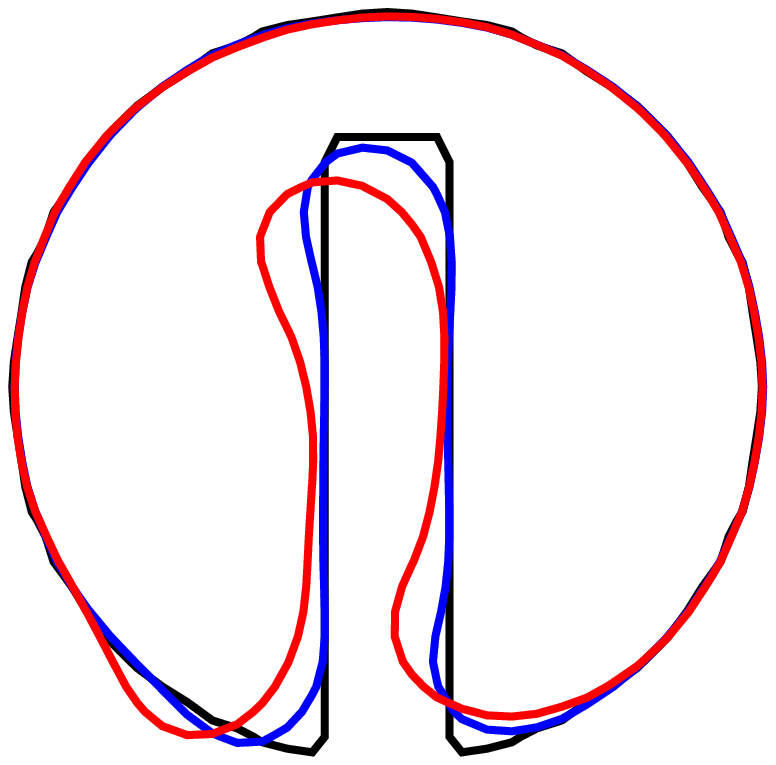}
  \includegraphics[width=0.24\textwidth]{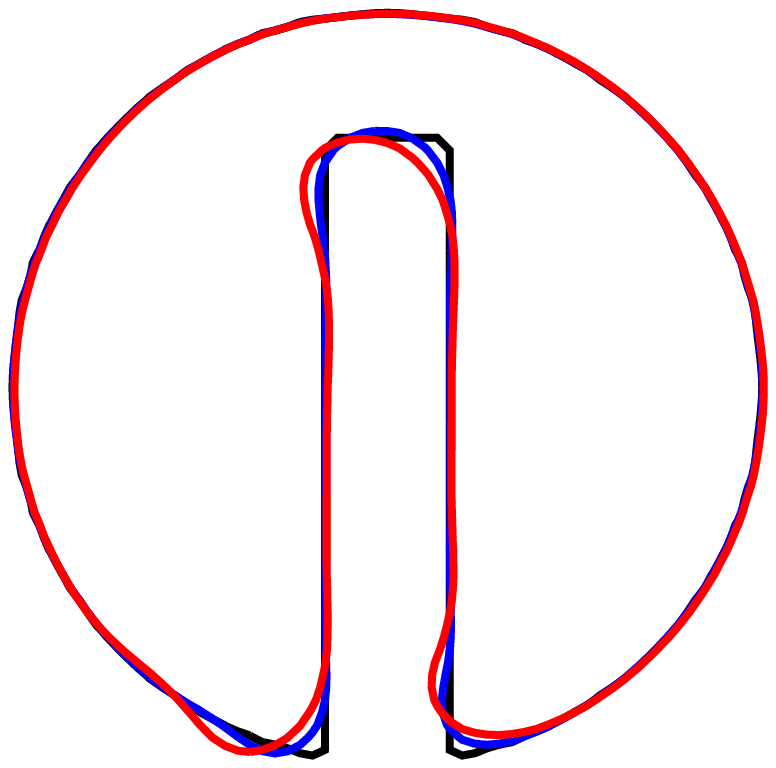}
  \includegraphics[width=0.24\textwidth]{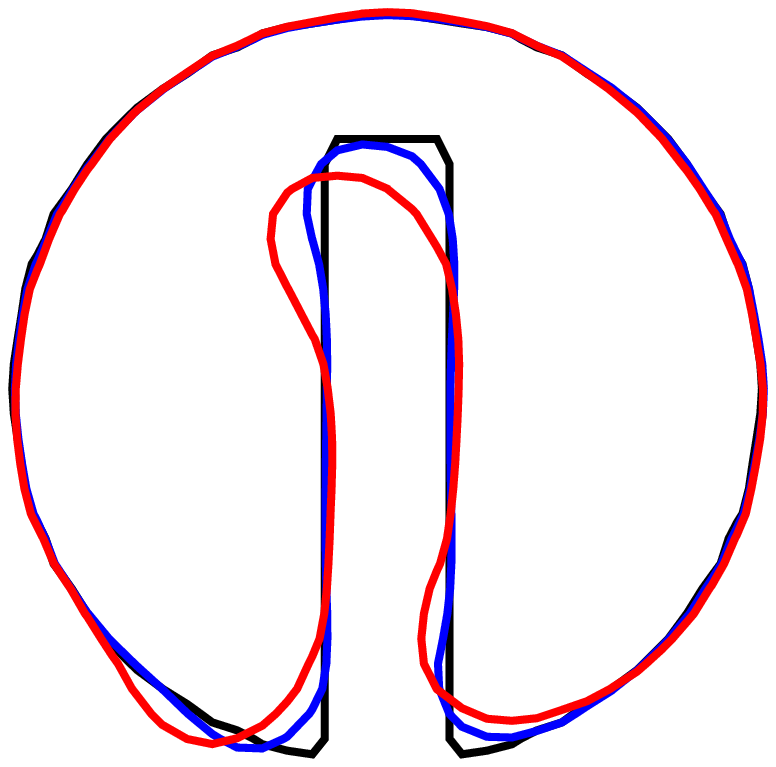}
  \includegraphics[width=0.24\textwidth]{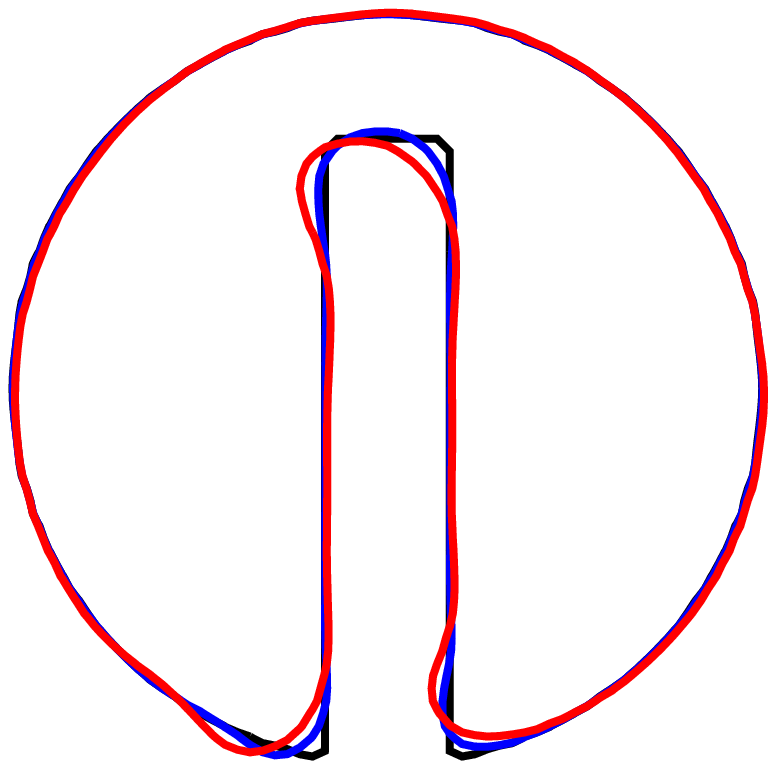}
 \caption{The $2D$ Zalesak's disk: simulation of a  slotted disc undergoing solid body rotation \citep{zalesak_1979a}. The black line denotes the exact initial solution, whereas blue and red the solutions obtained after one and five full revolutions. The two left figures are obtained with $\beta=1$ and the two right ones with $\beta=2$. In both sides, the results in the two panels are obtained with $33$ and $66$ grid points per diameter.}
  \label{fig:zal2d}
\end{figure}
\begin{figure}
  \centering
  \includegraphics[width=0.32\textwidth]{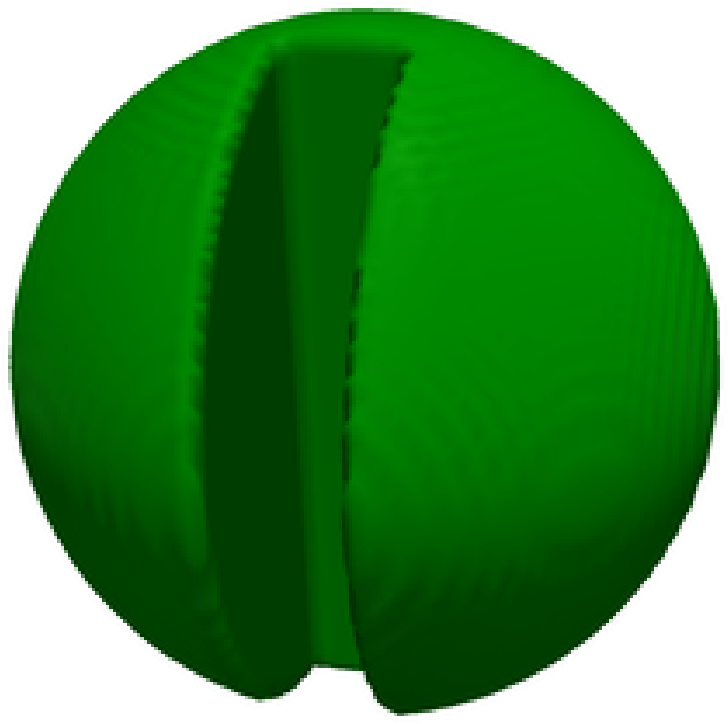}
  \includegraphics[width=0.32\textwidth]{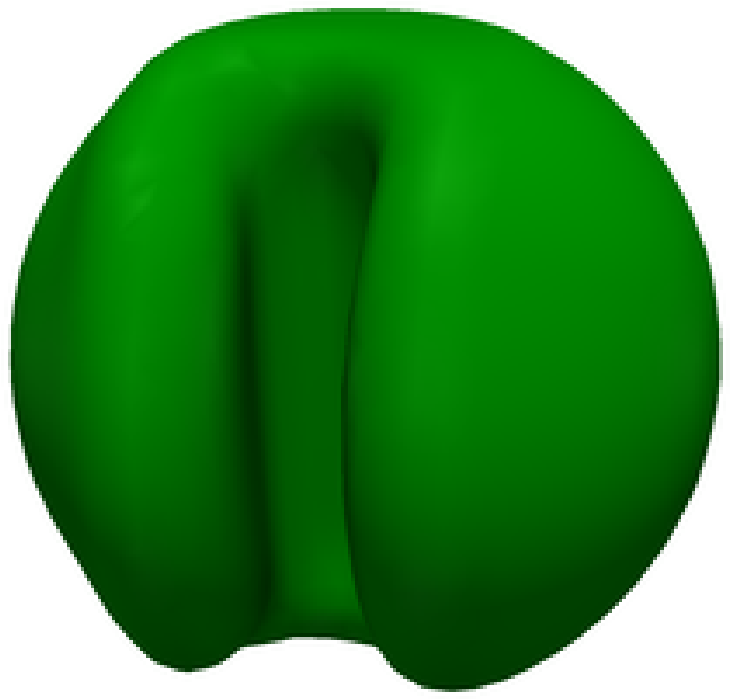}
  \includegraphics[width=0.32\textwidth]{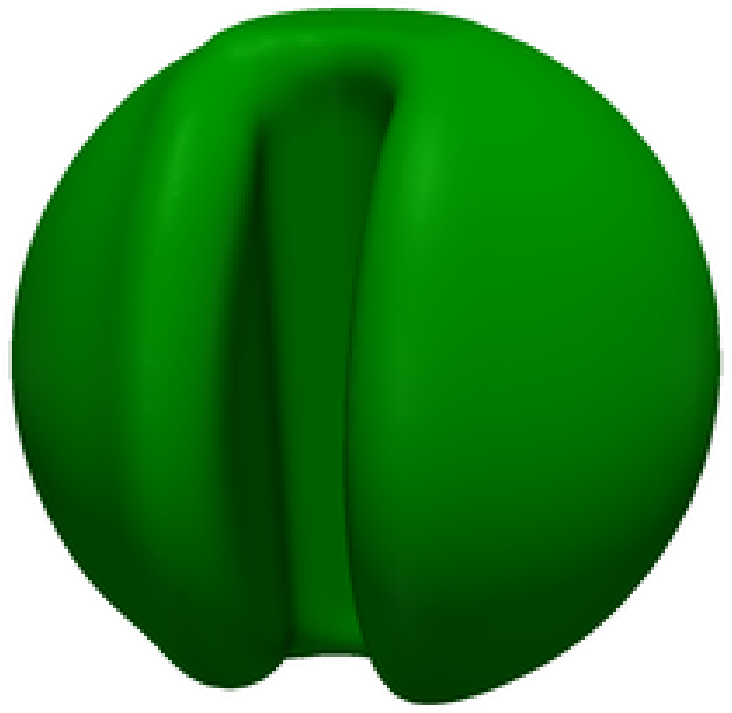}
 \caption{The $3D$ Zalesak's disk: the three panels represent the exact initial solution, and those after one full rotation with $33$ and $66$ grid points per diameter.}
  \label{fig:zal3d}
\end{figure}
The Zalesak's disk \citep{zalesak_1979a}, \ie a slotted disc undergoing solid body rotation, is a standard benchmark to validate numerical schemes for advection problems, since the initial shape should not deform under rigid body rotation. The set-up is the same described by \cite{ii_sugiyama_takeuchi_takagi_matsumoto_xiao_2012a} and the comparison of the initial shape (black line) and those after one (blue line) and five (red line) full rotations is shown in \figrefS{fig:zal2d}. We consider two grid resolutions here, $100$ and $200$ grid points per box size (being the disc of size $0.3$), and two different values of the sharpness parameter $\beta$: $1$ and $2$. The deformed shape of the disk shows an overall good agreement with the initial one, with the comparison deteriorating when more rotations are performed. Better agreement is found on the finer grid and this is further slightly improved in the case with $\beta=2$. As expected, the major differences are found on the sharp edges of the geometry, which are difficult to maintain undeformed. The test has been repeated in $3D$ and the results are shown in \figrefS{fig:zal3d}. Again, quite good agreement is found between the initial and final shapes, with the difference reducing with increasing resolution.

\begin{figure}
  \centering
  \includegraphics[width=0.24\textwidth]{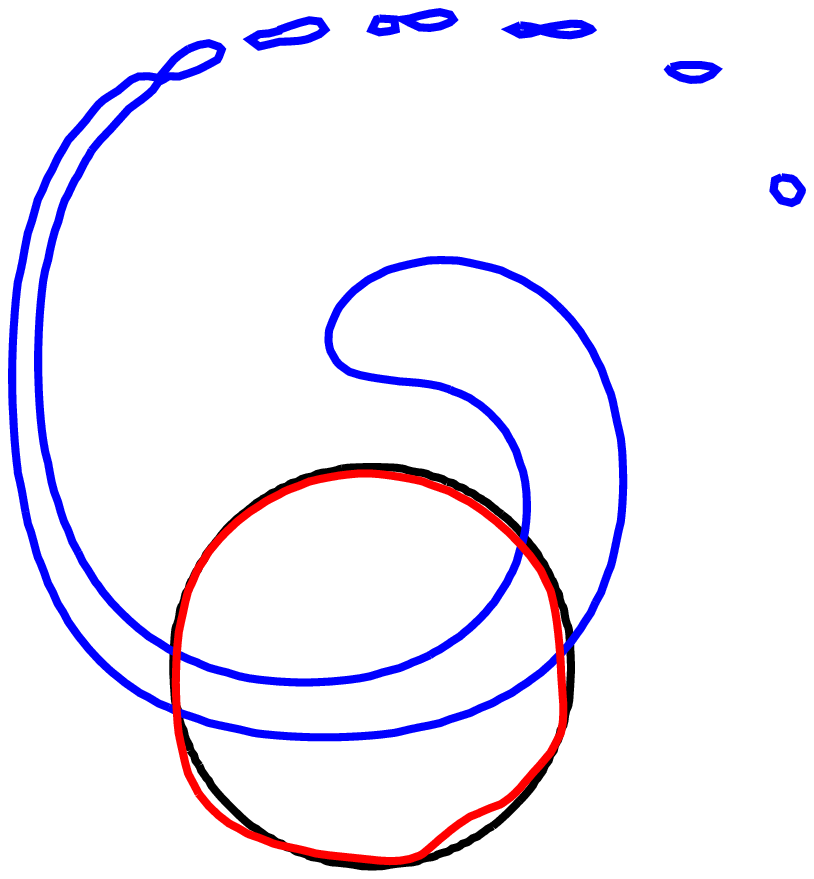}
  \includegraphics[width=0.24\textwidth]{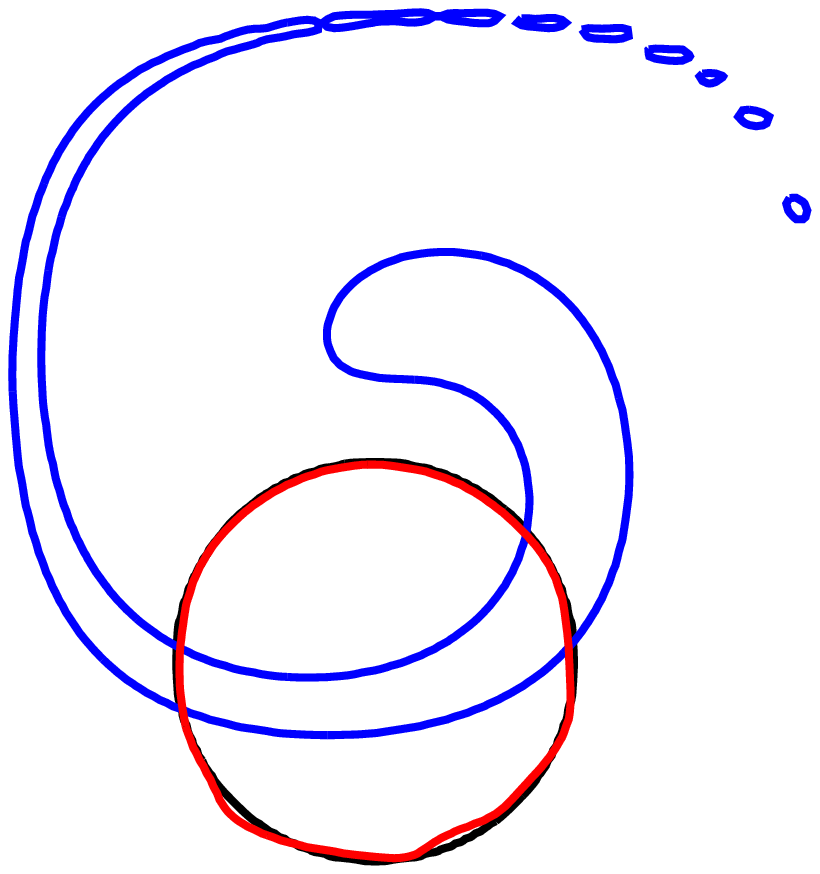}
  \includegraphics[width=0.24\textwidth]{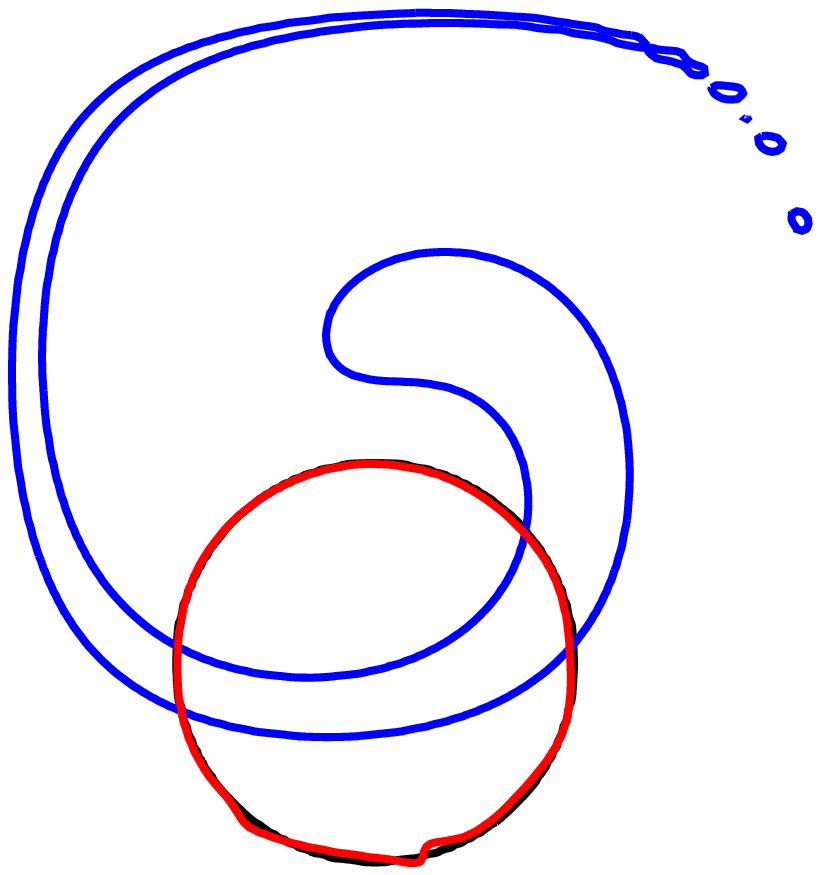}
 \caption{A deformed interface in a shearing flow: the black, blue and red lines denote the interface at $t=0$, $5\pi$ and $10\pi$, respectively. The three panels are obtained with $100$, $150$ and $200$ grid points per box side.}
  \label{fig:str2d}
\end{figure}
Next, we study a deformed interface in a shear flow in order to evaluate the capability of the method to capture a heavily deformed and stretched interface \citep{rudman_1997a}. We consider a circle of radius $0.2 \pi$, centered at $[0.5\pi,-0.2(\pi+1)]$ in a box of side $2\pi$ with the prescribed velocity field $u_1=\sin(x)\cos(y)$ and $u_2=-\cos(x)\sin(y)$ for $t<5\pi$, and with the reversed velocity for $t>5\pi$. First, the interface deforms due to the imposed velocity field reaching its maximum stretching at $t=5\pi$; after the flow reversal, the interface deformation starts reducing and at time $t=10\pi$ the interface goes back to its initial position. \figrefSC{fig:str2d} shows the interface shape at time $t=0$ (black line), at $t=5\pi$ (blue line) and at $t=10\pi$ (red line) for a mesh resolution of $100$ (left panel), $150$ (middle panel) and $200$ (right panel) grid points in each direction. We can observe that, the tail of the interface is broken in smaller segments at its maximum stretching ($t=5\pi$), and that the breakup reduces as the mesh is refined. In the reversed phase, all these interface segments merge and the shape is almost circular  at the final instant ($t=10\pi$), with some wiggles in the bottom part of the circle, yet reducing as the grid is refined.

We conclude this part on the numerical method performance by noting that the parasitic current often caused by numerical errors in the curvature estimation \citep{popinet_zaleski_1999a,torres_brackbill_2000a} are effectively suppressed in the present methodology by the continuous VOF distribution with the curved surface reconstruction, as shown by \cite{ii_sugiyama_takeuchi_takagi_matsumoto_xiao_2012a}.

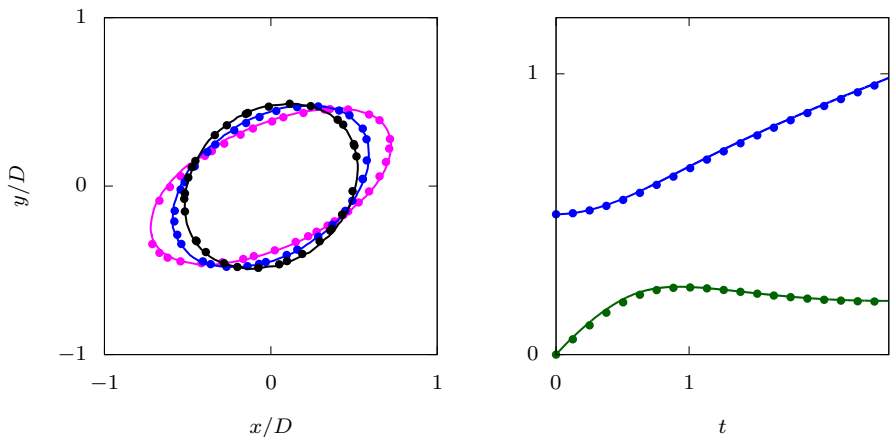
\begin{figure}
  \centering
  \input{shape}
  \input{bubble} \vspace{0.4cm}\\
 \caption{Validation of the multiphase solver, based on VOF with the multidimensional tangent of hyperbola interface capturing (MTHINC) method. a) Shape of a $3D$ deformable drop in shear flow for different Capillary numbers. The black, blue and magenta colors are used for $Ca=0.1$, $0.2$ and $0.3$, respectively. b) Center of mass position (blue) and rising speed (green) of a $2D$ rising droplet in a Newtonian fluid. In both panels, the solid lines are our results, while the points are results taken from the literature.}
  \label{fig:val}
\end{figure}
Finally we present two different validations of our code where we compare our results with those reported in the literature: first, we consider a $3D$ deformable drop in a simple shear flow, and then $2D$ and $3D$ rising droplets in a buoyancy-driven flow.

\paragraph*{Shear-driven flow -~~}
We consider a $3D$ unit spherical drop located at the center of a computational domain of size $8 \times 4 \times 8$, with a resolution of $16$ grid points for the drop diameter. The top and bottom boundaries move with opposite velocity $\pm U$, giving a shear rate $\dot{\gamma}=U/2$, while periodic boundary conditions are imposed in the streamwise $x$ and spanwise $z$ directions. The same density and viscosity are specified for the spherical drop and the surrounding fluid, the Reynolds number is fixed to $0.1$ and three Capillary numbers are studied: $0.1$, $0.2$ to $0.3$.  \figrefC[left]{fig:val} shows the steady-state deformed shapes obtained by our numerical simulations (solid lines), which are in excellent agreement with the numerical results reported by \citet{ii_sugiyama_takeuchi_takagi_matsumoto_xiao_2012a}, shown with points of the same color in the figure.

\paragraph*{Buoyancy-driven flow -~~}
The method is further validated for a $2D$ droplet rising in a fluid as studied numerically by \citet{hysing_turek_kuzmin_parolini_burman_ganesan_tobiska_2009a}. The domain is rectangular with size $1 \times 2$; no-slip boundary conditions are applied on the horizontal walls and free-slip boundary conditions on the sides. A round droplet with radius $R$ is initially placed at the centerline of the channel at a distance of $0.5$ from the bottom wall. The non-dimensional parameters governing the flow are the Reynolds number $Re=\rho_1U_g 2R/\mu_1$, the E${\rm\ddot{o}}$tv${\rm\ddot{o}}$s number $Eo=\rho_1 U_g^2 2R/\sigma$, the Capillary number $Ca=\mu_1 U_g/\sigma$ and the viscosity and density ratio $k_{\mu}=\mu_2/\mu_1$ and $k_{\rho}=\rho_2/\rho_1$, where we have chosen as reference length and velocity scales the drop diameter ($2R$) and the velocity $U_g=\sqrt{g L}$, being $g$ the gravitational constant. The simulation is performed with the parameters corresponding to the benchmark test case $1$ by \citet{hysing_turek_kuzmin_parolini_burman_ganesan_tobiska_2009a}, \ie $Re=35$, $Eo=10$, $Ca=0.2857$, $k_\mu=10$ and $k_\rho=10$. The comparison of the drop center of mass and rise velocity are reported in \figrefC[right]{fig:val}; the nearly perfect agreement between our results and those taken from the literature thus verifies again our implementation. Finally, we extend the comparison to a $3D$ droplet rising in a fluid as studied experimentally by \citet{legendre_daniel_guiraud_2005a}, ($Re=259$, $Eo=0.35$, $Ca=0.0025$, $k_\mu=1.56$ and $k_\rho=1.16$). From our simulation we obtain a rising velocity of $8.5$ cm/s, compared to $8.1$ cm/s found by \citet{legendre_daniel_guiraud_2005a}, with an error of $5\%$ well within the $10\%$ uncertainty in the velocity measurement reported by those authors.

\section{Results} \label{sec:result}
We consider the plane Couette flow of two Newtonian fluids separated by an interface with interfacial tension $\sigma$. The density and viscosity of the two fluids are assumed to be equal ($\rho_0$ and $\mu_0$), and the Reynolds number of the simulation is set equal to $\Rey = \rho_0 \dot{\gamma} r^2/\mu_0 = 0.1$, where $\dot{\gamma}$ is the reference shear rate, so that we can consider inertial effects negligible. We denote the solvent as fluid $1$ and the suspended phase as fluid $2$. The total volume fraction $\Phi$ is defined as the volume average of the local volume fraction $\phi$, \ie $\Phi=\bra{\bra{\phi}}$. Hereafter, the double $\bra{\bra{\cdot}}$ indicates time and volume average while the single $\bra{\cdot}$ average in time and in the homogeneous $x$ and $z$ directions. Four values of the total volume fraction $\Phi \approx 0.0016$, $0.1$, $0.2$, and $0.3$ are considered, together with three values of the interfacial tension $\sigma$, resulting in the Capillary numbers $\Ca = \muf \dot{\gamma}/\sigma =0.1$, $0.2$ and $0.4$. Note that, the Capillary numbers considered in this study are below the critical value for droplet breakup under shear flow in similar conditions \citep{cristini_guido_alfani_blawzdziewicz_loewenberg_2003a, caserta_simeone_guido_2008a}; indeed, \citet{caserta_simeone_guido_2008a} and \citet{caserta_guido_2012a} found that this system is mainly modified by deformation and coalescence, thus these effects will be investigated in our simulations. The numerical domain is a rectangular box of size $16r \times 10r \times 16r$ in the $x$, $y$, and $z$ directions, discretised on a Cartesian uniform mesh with $16$ grid points per radius $r$. No-slip boundary conditions are imposed on the solid walls, while periodic boundary conditions are enforced in the homogeneous $x$ and $z$ directions (see \figrefS{fig:sketch}). All simulations start with a stationary flow and a random distribution of spherical droplets of fluid $2$ with unit radius across the domain. Note that, the general set-up and most of the parameters used in this study are chosen as in \citet{picano_breugem_mitra_brandt_2013a} and \citet{rosti_brandt_2018a} where suspensions of rigid and deformable spherical particles are examined. The independence of the results from the grid resolution and domain size were tested by performing two additional simulations for the case with high volume fraction and Capillary number, \ie $\Phi=0.3$ and $\Ca=0.4$: i) a simulation with double grid points in all the directions; ii) a simulation with double domain size in the homogeneous directions $x$ and $z$. The difference in the results was found to be less than $2\%$.

\begin{figure}
  \centering
  \input{vel}
  \input{vof}\\
  \vspace{0.4cm}
  \caption{(left) Mean solvent streamwise velocity profile $u^{f1}$, normalized with the wall velocity $V_w$, and (right) average local volume fraction distribution, $\phi$, along the wall-normal direction $y$. Different colors are used to represent different volume fractions $\Phi$: $10\%$ (blue), $20\%$ (orange) and $30\%$ (red).}
  \label{fig:velVof}
\end{figure}
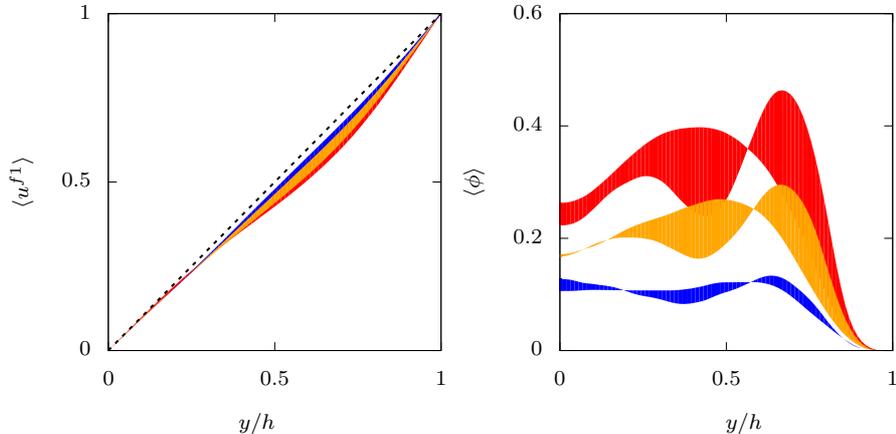
We start the analysis of the two-fluid flow by showing the mean streamwise velocity profile of the solvent phase $u^{f1}$ in the left panel of \figrefS{fig:velVof}. The blue, orange and red colors are used to distinguish the three different volume fractions $\Phi=0.1$, $0.2$ and $0.3$, respectively; the shaded area with the same color represents the spread of the data due to the different Capillary numbers studied in this work. In the Newtonian case, shown with a dashed black line, the velocity profile is linear, decreasing from $V_w$ at the wall, due to the no-slip condition, to zero at the center line, for symmetry. The wall-normal profile is not straight in the emulsion and the velocity decreases faster than in the Newtonian case close to wall, \ie the wall-normal derivative of the velocity profile at the wall increases; also, the profile shows a local minimum around $y\approx0.75h$. These differences are enhanced for high values of volume fractions and Capillary numbers, and are strongly related to the distribution of the dispersed phase across the channel, reported
in terms of average local volume fraction of the suspended fluid $\bra{\phi}$ in the right panel of the same figure. Indeed, the suspended fluid has a non uniform distribution in the wall-normal direction, with a strong peak in the concentration around $y\approx0.75h$, (the position of the local minimum of velocity), which moves towards the wall for increasing volume fractions. From the figure we can also observe a significant variation with the Capillary number as the volume fraction increases; this result differs from what reported by \citet{rosti_brandt_2018a} for deformable particles, where only minor variations where observed in the volume fraction distribution for different Capillary numbers.

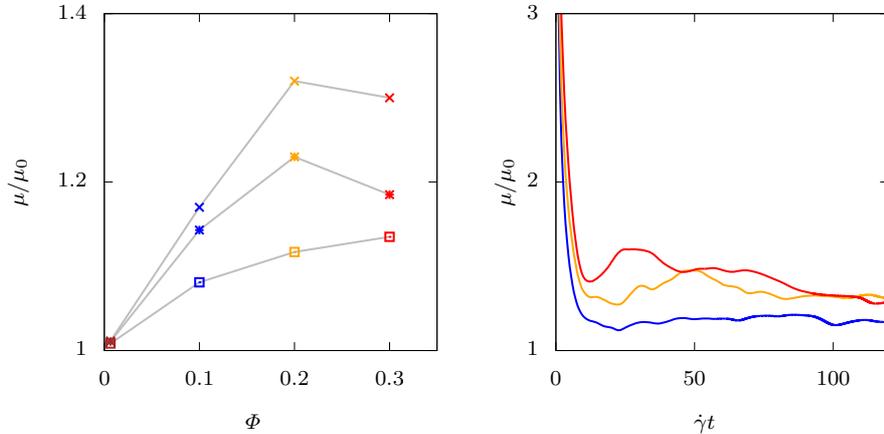
\begin{figure}
  \centering
  \input{viscF}
  \input{viscT}\\
  \vspace{0.4cm}
  \caption{(left) The effective viscosity $\mu/\mu_0$ as a function of the volume fraction $\Phi$ for different values of the volume fraction, $\Phi \approx 0.0016$ (brown), $0.1$ (blue), $0.2$ (orange), and $0.3$ (red), and of Capillary number $\Ca=0.1~(\times)$, $0.2~(\ast)$, and $0.4~(\boxdot)$. (right) Time evolution of the effective viscosity $\mu/\mu_0$ for volume fractions $\Phi = 0.1$ (blue), $0.2$ (orange), and $0.3$ (red) and for Capillary number $\Ca=0.1$.}
  \label{fig:viscFG}
\end{figure}

The wall-normal derivative of the streamwise velocity at the wall can be used to estimate the effective viscosity of the non-Newtonian emulsion made of the two Newtonian fluids. We define the effective 
viscosity $\mu$, normalized by the reference value $\mu_0$, as 
\begin{equation} \label{eq:visc}
\frac{\mu}{\mu_0}=\frac{\bra{\bra{\sigma_{12}^{\rm w}}}}{\mu_0 \dot{\gamma}}.
\end{equation}

\figrefC[left]{fig:viscFG} shows the effective viscosity $\mu$ as a function of the total volume fraction $\Phi$ for all the Capillary number $\Ca$ considered. We observe that the effective viscosity decreases with the Capillary number $\Ca$ and exhibits a non-monotonic behavior with respect to the volume fraction: in particular, it first increases with the volume fraction up to $\Phi=20\%$ and then decreases. These data are obtained by time-averaging the results over a time of approximately $40\dot{\gamma}^{-1}$ after a long transient of approximately $80\dot{\gamma}^{-1}$. Indeed, the system is reaching a quasi statistically steady state after this interval for all the cases we have investigated, as shown in \figref[right]{fig:viscFG} where we report the time history of the effective viscosity $\mu$ for three different volume fractions ($\Phi=0.1$, $0.2$ and $0.3$) and Capillary number $Ca=0.1$. Note that, we have defined as quasi statistically steady state regime a condition where the global properties of the system (\eg the effective viscosity) change by less than $3\%$ over a time of $20\dot{\gamma}^{-1}$.

Next, we test whether an expression similar to that used for rigid and deformable objects can be used also for emulsions. To this end, we fit the effective viscosity of our simulations with the relation proposed by \citet{batchelor_green_1972a}, \ie a second order extension of \citet{einstein_1956a} formula,
\begin{equation} \label{eq:batchelor}
\dfrac{\mu}{\muf}=1+\left[\mu\right]\Phi+B_{BG} \Phi^2,
\end{equation}
where $\left[\mu\right]$ is the intrinsic viscosity (equal to $5/2$ for rigid dilute particles) and $B_{BG}$ is a coefficient equal to $7.6$ for non-Brownian spheres. In the two-fluids system studied here, we measure the intrinsic viscosity $\left[\mu\right]$ from the simulations with a single droplet in the computational domain (corresponding to the case $\Phi \approx 0.0016$), \ie $\left[\mu\right] \approx \left( \mu - \mu_0 \right)/ \left( \mu_0 \Phi \right)$. The coefficient $B_{BG}$ is kept as a fitting parameter, whose values are reported in \tabref{tab:fit}, together with the intrinsic viscosity extracted from our simulations. We see that the second-order expression is applicable for values of the volume fraction up to $\Phi=0.3$ in the two-fluid system considered here, provided that both $\left[\mu\right]$ and $B_{BG}$ are modified to take into account the droplet deformation, \ie they are assumed as functions of the Capillary number $\Ca$. For rigid particle suspensions, this second-order relation is usually inaccurate for $\Phi \gtrsim 0.15$, as the viscosity increases faster than a second order polynomial \citep{stickel_powell_2005a}; however, it has been shown recently to apply up to $\Phi \approx 0.33$ for suspensions of deformable particles \citep{rosti_brandt_2018a}.
\begin{table}
\centering
\setlength{\tabcolsep}{5pt}
\begin{tabular}{l|cc}
$\Ca$	&	$\left[ \mu \right]$	&	$B_{BG}$ 			\\ \hline
$0.1$	&	$1.7538$					&	$-2.2078$		\\
$0.2$	&	$1.6615$					&	$-3.4116$	\\
$0.4$	&	$1.4769$					&	$-3.6253.$	\\
\end{tabular}
\caption{The fitting parameter $B_{BG}$ in \equref{eq:batchelor} used for the curves in \figrefS{fig:viscFG} and the intrinsic viscosity $\left[\mu\right]$ computed from our simulations.}
\label{tab:fit}
\end{table}
From the data in the table, we observe that in the case of emulsions both the intrinsic viscosity $\left[\mu\right]$ and the second-order coefficient $B_{BG}$ are decreasing with the Capillary number $\Ca$. Also, $B_{BG}$ is negative for all the Capillary numbers $\Ca$: the function $\mu = \mu \left( \Phi \right)$ has a maximum. This can be explained by considering the limiting behaviors for $\Phi=0$ and $1$; indeed, the limit for $\Phi \rightarrow 0$ and $\Phi \rightarrow 1$ is $\mu = \mu_0$, \ie we recover the fluid viscosity, thus the function $\mu = \mu \left( \Phi \right)$ must show a maximum for $0<\Phi<1$. These results are in good agreement with the experiments by \citet{caserta_simeone_guido_2006a, caserta_simeone_guido_2008a}.

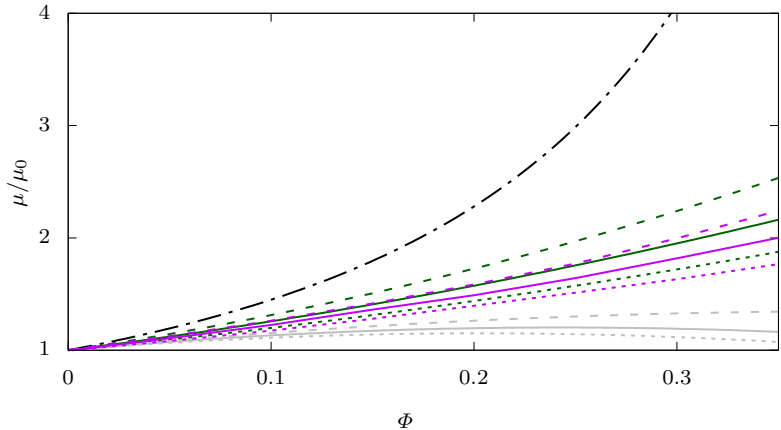
\begin{figure}
  \centering
  \input{viscALL}\\
  \vspace{0.55cm}
  \caption{The effective viscosity $\mu/\mu_0$ as a function of the volume fraction $\Phi$ for three Capillary numbers $\Ca$: $0.1$ (dashed line), $0.2$ (solid line) and $0.4$ (dotted line). The grey lines are the present results for a two-fluids system, the green and purple are results taken from the literature for deformable elastic particles \citep{rosti_brandt_2018a} and capsules \citep{matsunaga_imai_yamaguchi_ishikawa_2016a}, respectively, while the black dash-dotetd line represent the rigid particle limit \citep{picano_breugem_brandt_2015a}.}
  \label{fig:viscALL}
\end{figure}
Finally, we compare in \figrefS{fig:viscALL} the effective viscosity $\mu$ of the present two-fluids system with the results taken from the literature at similar volume fraction and Capillary number. In particular, the present results are shown in the figure with grey lines, while green lines represent the case of deformable hyper-elastic particles \citep{rosti_brandt_2018a}, the purple lines indicate the case of deformable capsules \citep{matsunaga_imai_yamaguchi_ishikawa_2016a} and the black line is the results for rigid particles \citep{picano_breugem_brandt_2015a}. In the deformable cases reported (two-fluids system, capsules and elastic particles), results pertaining three different Capillary numbers are shown in the figure: $\Ca=0.1$ (dashed line), $0.2$ (solid line) and $0.4$ (dotted line). We clearly note that, all the deformable cases show smaller effective viscosity than in the presence of rigid particles, with the difference increasing with the volume fraction $\Phi$ and the Capillary number $\Ca$. Also, the present results are those with lower effective viscosity, followed by capsules and finally by deformable particles whose results are the closest to the one for rigid particles. The reduction of the effective viscosity with respect to the case of rigid spheres indicates a reduced influence of the suspended phase on the flow of the carrier phase. This is due to the deformability which allows the intrusions to attain a shape which is able to reduce the obstruction to the flow and adapts to it.

\begin{figure}
  \centering
  \includegraphics[width=0.322\textwidth]{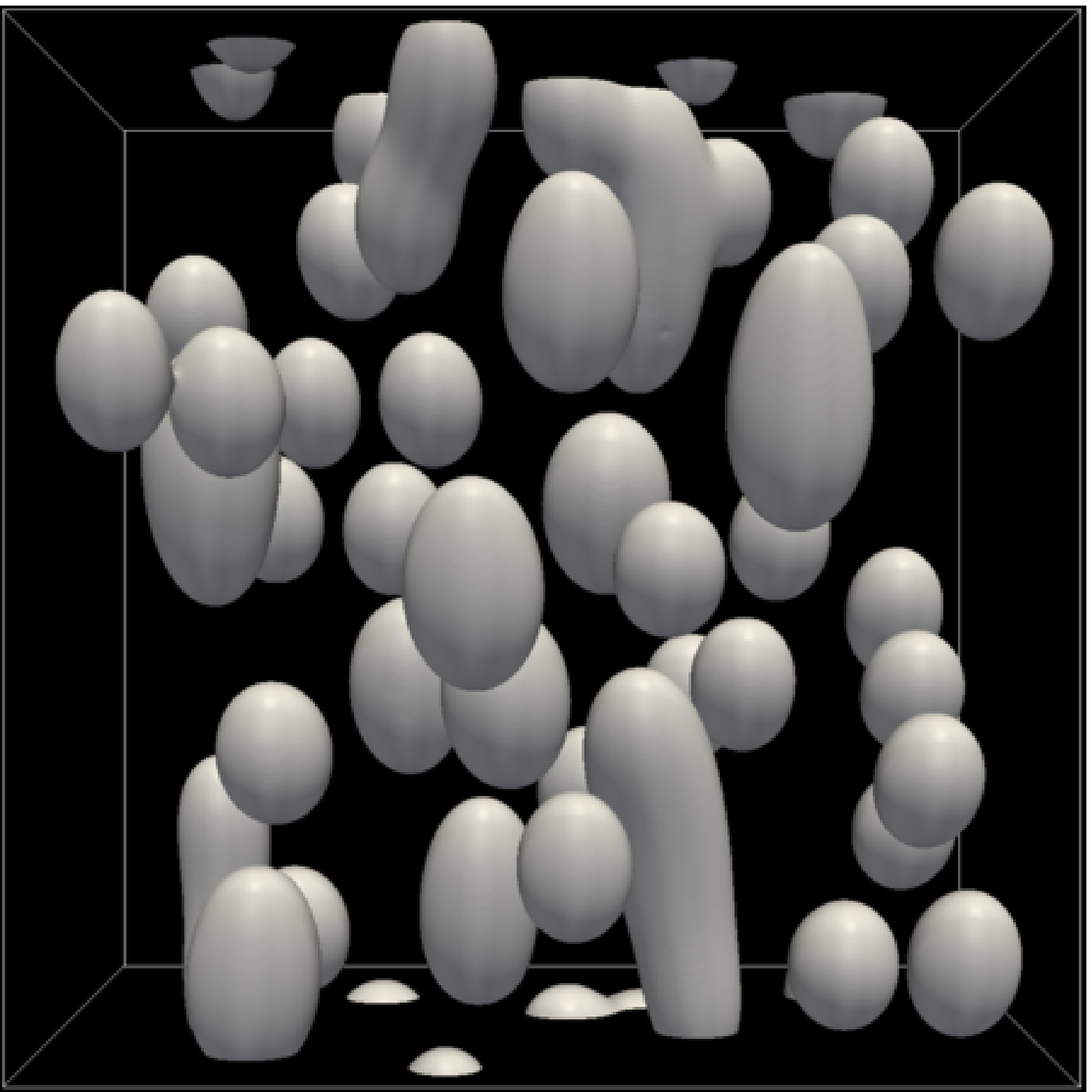}
  \includegraphics[width=0.322\textwidth]{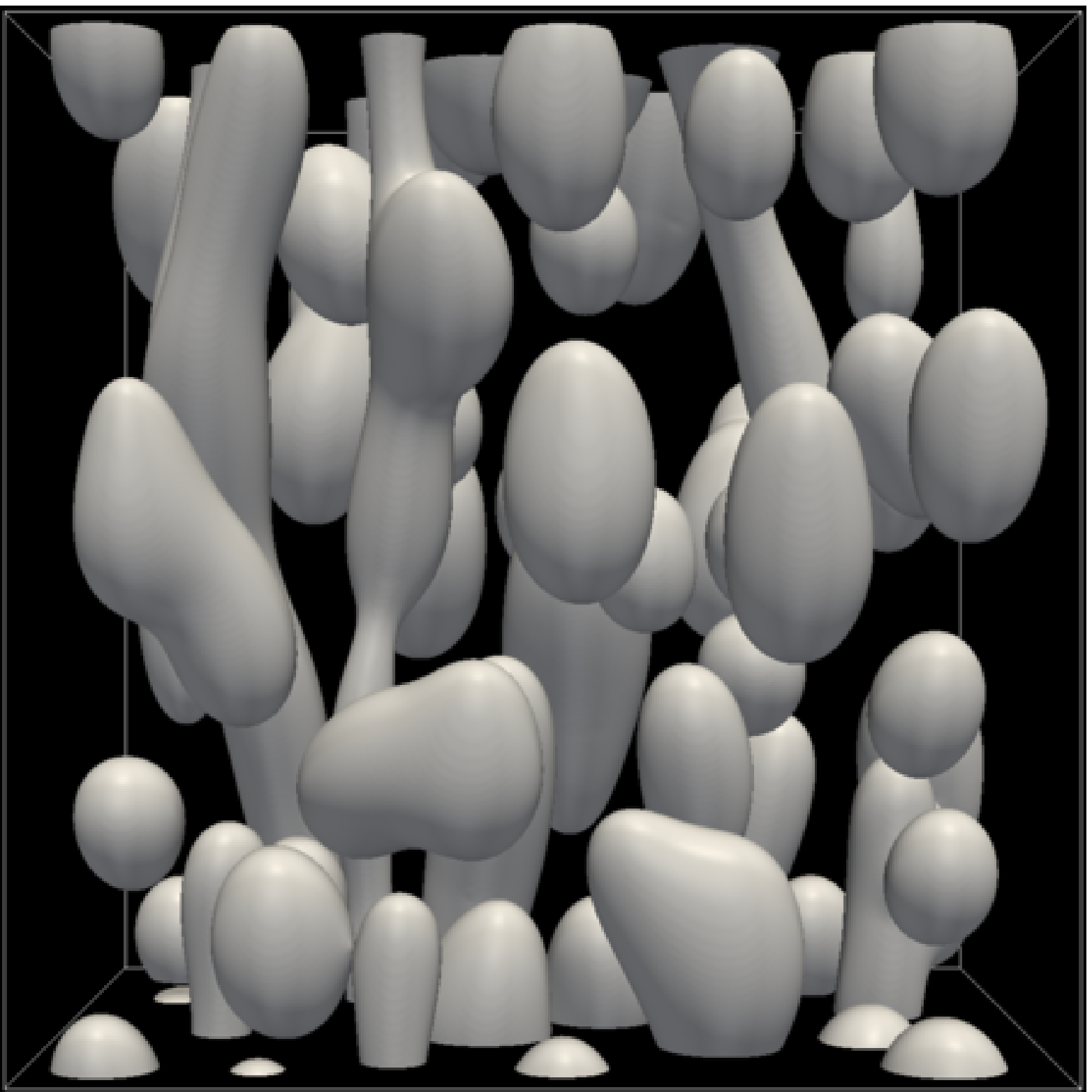}
  \includegraphics[width=0.322\textwidth]{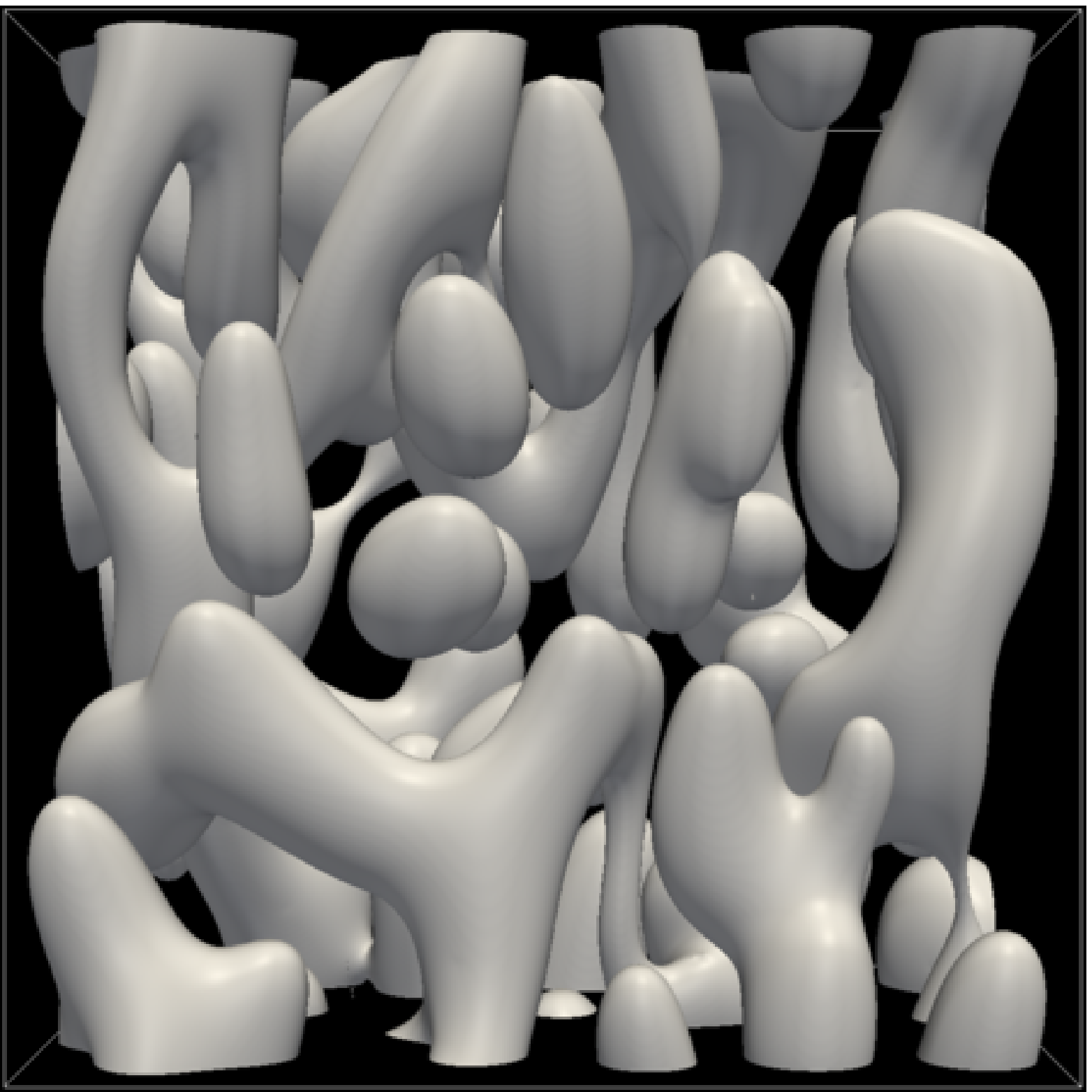}\\ \vspace{0.07cm}
  \includegraphics[width=0.322\textwidth]{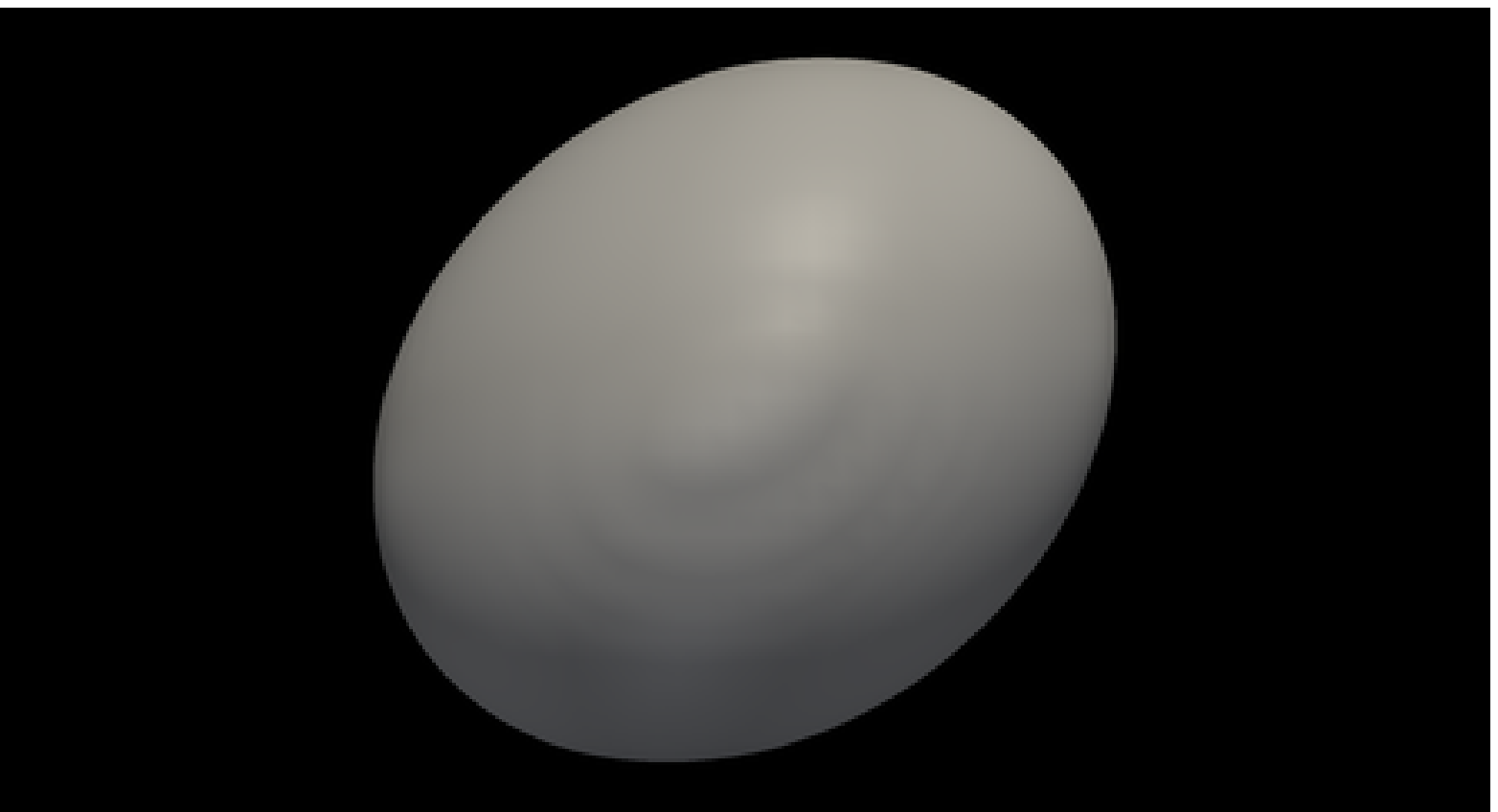}
  \includegraphics[width=0.322\textwidth]{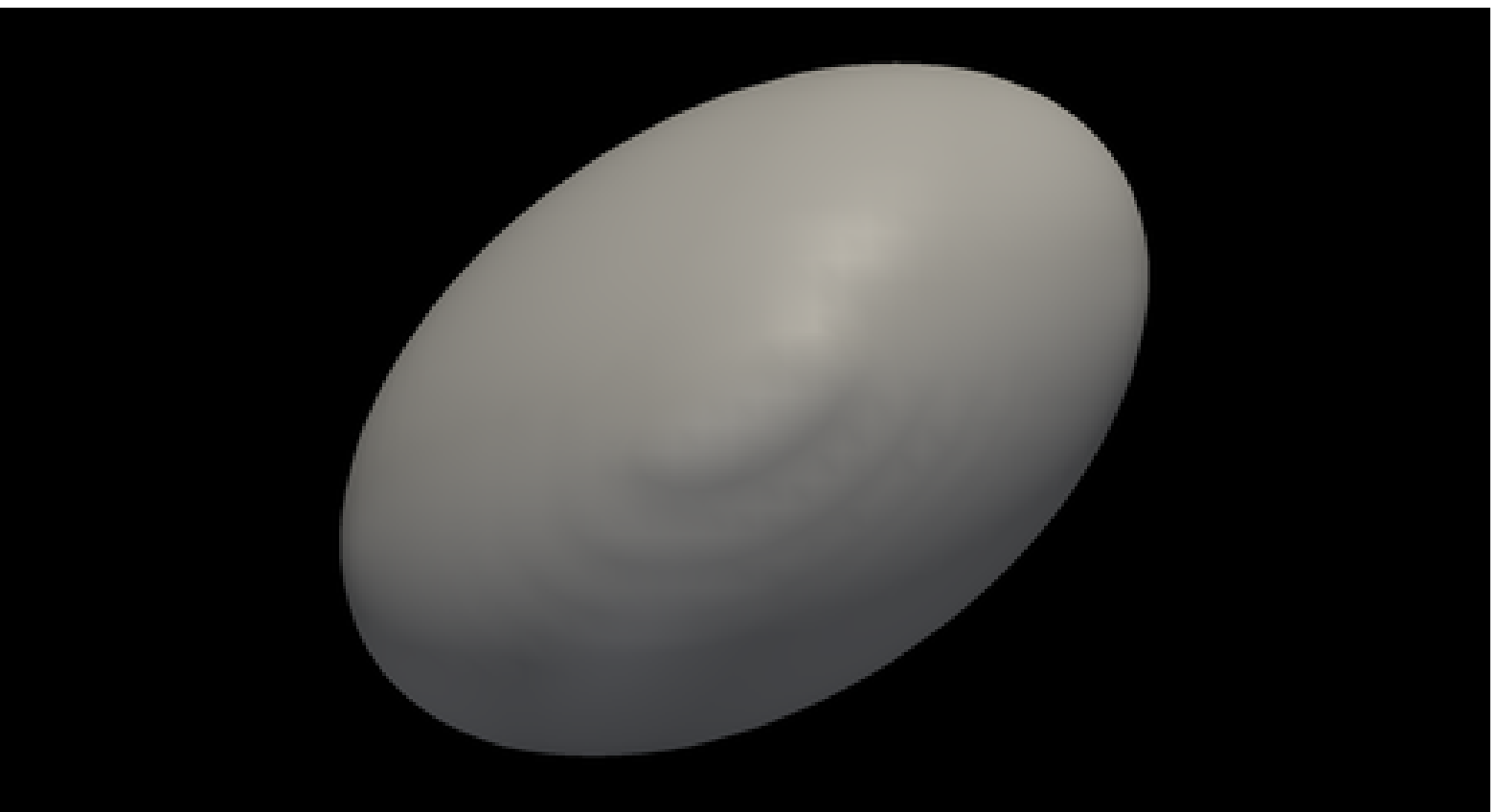}
  \includegraphics[width=0.322\textwidth]{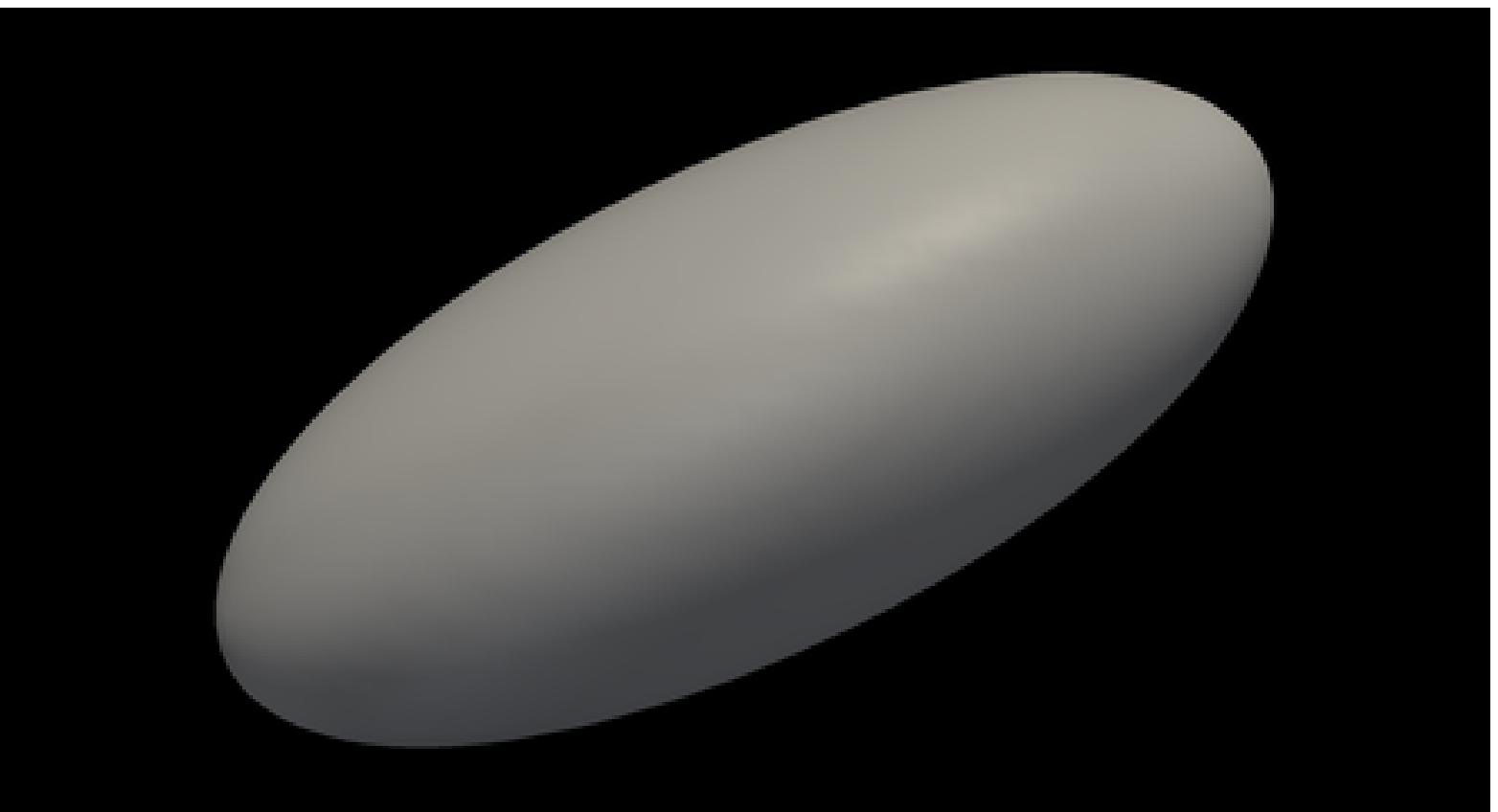}
  \caption{(top) Instantaneous shape of the suspension at $\Ca=0.2$ for three different volume fractions: $\Phi=0.1$, $0.2$, and $0.3$. (bottom) Shape of the particle in the dilute case, $\Phi=0.0016$, for increasing Capillary numbers $\Ca=0.1$, $0.2$ and $0.4$.}
  \label{fig:shape}
\end{figure}

We now study the behavior of the interface in the emulsions under investigations here. \figrefSC{fig:shape} displays instantaneous configurations of the interface between the two fluids: the top row reports cases at $\Ca=0.2$ and increasing $\Phi$, \ie $0.1$, $0.2$ and $0.3$, while the bottom row is for the dilute suspension $\Phi=0.0016$ and increasing $\Ca$, \ie $0.1$, $0.2$ and $0.4$. We observe that, in the dilute case, the shape progressively changes from a sphere to an ellipsoid as the Capillary number $\Ca$ increases. However, in the case of an emulsion, the interface shape rapidly degenerates from that of an ellipsoid due to the merging process: indeed, we can observe the formation of elongated structures spanning large portions of the domain both in the streamwise and spanwise directions. To quantify the deformation, it is common in the literature to evaluate the so-called Taylor parameter
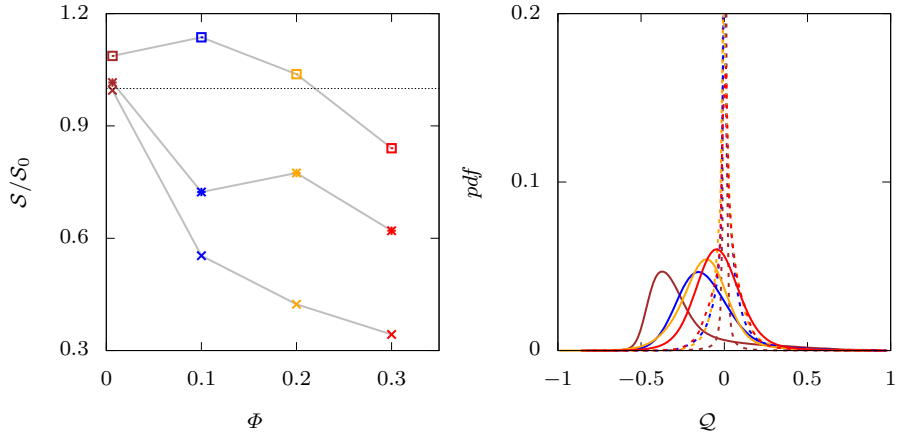
\begin{figure}
  \centering
  \input{SF}
  \input{pdf}\\
  \vspace{0.55cm}
  \caption{(left) Normalized mean interface surface area $\mathcal{S}$ as a function of the volume fraction $\Phi$ for different Capillary numbers $\Ca$. The symbol and color scheme is the same as in \figrefS{fig:viscFG}. (right) Probability density function of the flow topology parameter $\mathcal{Q}$ for all the volume fractions $\Phi$ and for a fixed Capillary number $\Ca=0.2$. The color scheme is the same as in \figrefS{fig:viscFG}, and the dashed and solid lines are used for fluid $1$ and $2$, respectively.}
  \label{fig:def}
\end{figure}
\begin{equation} \label{eq:Taylor}
\mathcal{T}=\frac{a-b}{a+b},
\end{equation}
where $a$ and $b$ are the semi-major and semi-minor axis of the inscribed ellipse passing through the center of the droplet in the $x$-$y$ plane. However, due to the merging, only few interface shapes are truly ellipsoidal, and thus this measure is not appropriate for the present case. An alternative indicator of the deformation and merging of the droplets is reported in the left panel of \figrefS{fig:def}, where the mean interface surface area $\mathcal{S}$, normalized by its initial value $\mathcal{S}_0$, is depicted as a function of the total volume fraction $\Phi$. We observe that $\mathcal{S}$ decreases (although not monotonically) with the volume fraction $\Phi$ and increases with the Capillary number $\Ca$. In the dilute case (brown symbols) the surface is always greater than the initial spherical shape, as expected being the sphere the shape with lowest surface area to volume ratio. As $\Ca$ increases and the average shape changes going from a sphere to an ellipsoid (see the bottom panels in \figrefS{fig:shape}) the surface area increases. Thus, at a fixed Capillary number and increasing  volume fraction, the total surface area reduces due to the merging. We can therefore conclude the following: \textit{i)} for the low Capillary number cases ($\Ca=0.1$, displayed with the symbol $\times$) the merging is the dominant effect because the droplets are only slightly deforming and most of the deviations from the initial configuration are due to the merging; \textit{ii)} for the high Capillary number cases ($\Ca=0.4$, shown by $\boxdot$) the deformation is dominant at low volume fractions when the surface area is increasing, but eventually, as $\Phi$ increases, the merging reduces the total surface area; \textit{iii)} finally, the cases with intermediate Capillary number ($\Ca=0.2$, indicated with $\ast$) present a mixed behavior, although at high $\Phi$ the merging always prevails, reducing the total area.

In order to evaluate the different flow behaviors in the two fluids, we compute the so called flow topology parameter $\mathcal{Q}$, successfully used by \citet{de-vita_rosti_izbassarov_duffo_tammisola_hormozi_brandt_2018a} among others. The flow topology parameter is defined as
\begin{equation} \label{eq:q}
\mathcal{Q} = \frac{D^2 - \Omega ^2}{D^2 + \Omega^2},
\end{equation}
where $D^2 = D_{ij}D_{ji}$ and $\Omega^2 = \Omega_{ij}\Omega_{ji}$, being $\Omega_{ij}$ the rate of rotation tensor, $\Omega_{ij} = (\partial u_i/\partial x_j - \partial u_j/\partial x_i)/2$. When $\mathcal{Q} = -1$ the flow is purely rotational, regions with $\mathcal{Q} = 0$ represent pure shear flow and those with $\mathcal{Q} = 1$ elongational flow. The distribution of the flow topology parameter for the cases with $\Ca=0.2$ and different volume fractions $\Phi$ is reported in \figrefC[right]{fig:def}. In particular, we show the probability density function (pdf) of $\mathcal{Q}$ in the two liquid phases separately. We observe that in the fluid $1$ the flow is mostly a shear flow, as demonstrated by a sharp peak at $\mathcal{Q} = 0$. On the other hand, the flow of fluid $2$ shows a broad peak for $\mathcal{Q} < 0$, meaning that the flow is more rotational. However, this peak displaces towards $\mathcal{Q} = 0$ as the volume fraction is increased. This is caused by the increased merging at high $\Phi$, which generates large structures (see \figrefS{fig:shape}) which spans larger and larger area of the domain.

\begin{figure}
  \centering
  \input{tauG11}
  \input{tauhist1}\\
  \vspace{0.8cm}  
  \input{tauG22}
  \input{tauhist2}\\
  \vspace{0.8cm}  
  \input{tauG33}
  \input{tauhist3}\\
  \vspace{0.4cm}
  \caption{Decomposition of the total shear stress $\bra{\sigma_{12}}$ in its contributions (see \equref{eq:stressdecomp}), normalized by the total wall value $\bra{\sigma_{12}^{\rm w}}$. The figures in the left column show the profile of the shear stress components as a function of the wall-normal distance for the case with $\Ca=0.2$ while those in the right column the percentage distribution of the shear stress contribution as a function of the Capillary number $\Ca$. The top, middle and bottom row indicate the three volume fraction $\Phi=0.1$, $0.2$ and $0.3$, respectively. The shear stress contribution due to the interfacial tension, viscous stress of fluid $1$ and viscous stress of fluid $2$ are shown with the dash-dotted $-\cdot-$, dashed $-~-$ and solid --- lines in the right column and with the grey, blue and green bars in the histograms in the right column.}
  \label{fig:tauij}
\end{figure}
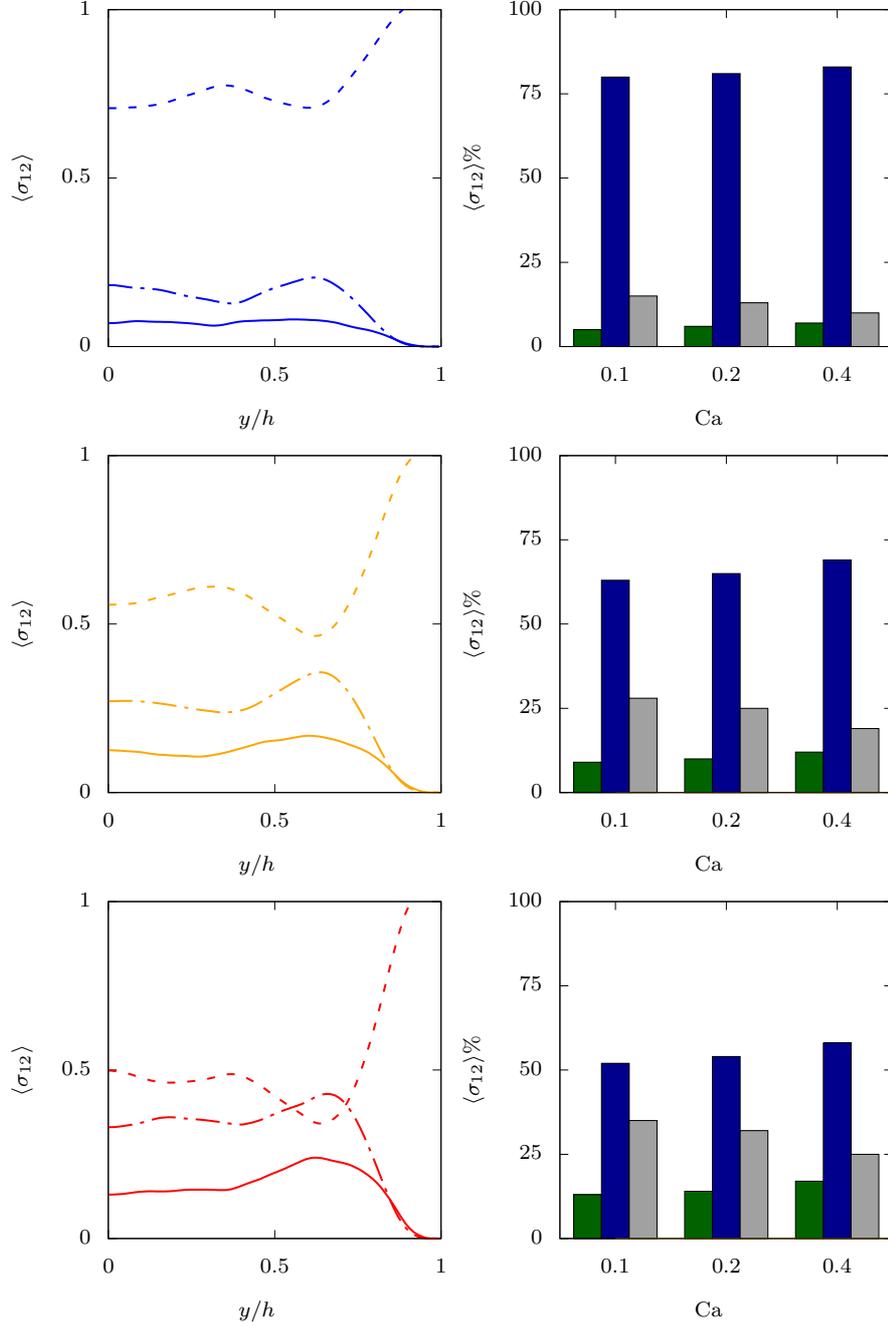

To understand the rheological behavior of the emulsion, we consider the different contributions to the momentum transfer. The streamwise component of the mean momentum equations for a Couette flow with streamwise and spanwise homogeneous direction reads \citep{pope_2001a}
\begin{equation} \label{eq:der1}
\muf \frac{d^2\bra{u}}{dy^2} - \rho \frac{d \bra{u'v'}}{dy} + \bra{f_{1}} = 0,
\end{equation}
where the $'$ indicates the fluctuation with respect to the mean. Integrating the equation in the wall-normal direction $y$ we obtain
\begin{equation} \label{eq:der2}
\frac{d\bra{\sigma_{12}}}{dy} = 0,
\end{equation}
where we have defined the total shear stress $\bra{\sigma_{12}}$ as the sum of the viscous and Reynolds stresses, with the addition of a contribution from the interfacial tension $f_i$, \ie
\begin{equation} \label{eq:der3}
\bra{\sigma_{12}} = \muf \frac{d\bra{u}}{dy} - \rho \bra{u'v'} + \int_0^y \bra{f_{1}}.
\end{equation}
Note that, although the integral of the interfacial tension over a full droplet is null, the wall-normal profile of its contribution is not, and it must thus be accounted for. Finally, the viscous and Reynolds stresses can be further decomposed into the contribution of the two fluids, see \equref{eq:phi-stress}, giving the following expression:
\begin{multline} \label{eq:stressdecomp}
\bra{\sigma_{12}} = \underbrace{\left( \muf \frac{d\bra{u}}{dy} - \rho \bra{u'v'} \right)}_{\rm fluid~1} + \underbrace{ \left( \muf \frac{d\bra{u}}{dy} - \rho \bra{u'v'} \right)}_{\rm fluid~2} + \underbrace{\int_0^y~\bra{f_{1}}~dy}_{\rm interfacial~tension},
\end{multline}
where the averages in the first two brackets are performed separately in the two phases. Each of these contribution have been averaged in time and in the homogeneous directions and displayed in \figref[left]{fig:tauij} as function of the wall-normal distance $y$ for the cases with $\Ca=0.2$ and volume fractions $\Phi=0.1$, $0.2$ and $0.3$ (top, middle and bottom panel in the left column). Note that, the stresses are normalized by the total wall value, thus they vary between $0$ and $1$ and that the Reynolds stress contributions are not shown being negligible in the inertialess limit. 

The viscous stress of fluid $1$ is the only not null component at the wall, due to the fact that the fluid $2$ has null concentration there (see \figref[right]{fig:velVof}). At the lowest volume fraction shown here the fluid $1$ viscous stress is the dominant contribution, being responsible for more than $70\%$ of the total stress at each wall-normal location. It has a minimum value around $y\approx0.75h$ corresponding to the location of maximum concentration of fluid $2$ as previously discussed. As the volume fraction is increased, the fluid $1$ viscous stress becomes smaller and smaller; this is compensated by an increase in the fluid $2$ viscous stress contribution. Moreover, we observe that the interfacial tension term is almost uniform across the channel, and that its contribution increases with the volume fraction. In \figref[right]{fig:tauij} we summarize the stress budgets analysis by showing the volume-averaged percentage contribution of all the non-zero components of the total shear stress, \ie fluid $1$ viscous stress (blue), fluid $2$ viscous stress (green), and interfacial tension (grey). Each panel shows how the stress balance change with the Capillary number, and each panel corresponds to a different volume fraction ($\Phi=0.1$: top panel; $0.2$: middle panel; $0.3$: bottom panel). For every volume fraction, we observe that the percentage contribution of the viscous stress of the solvent and suspended fluids increases with the Capillary number; on the other hand, the contribution of the interfacial tension slightly decreases with the Capillary number (see \figref[left]{fig:def}). These results are similar to the one reported for hyper-elastic particles by \citet{rosti_brandt_2018a} who also found an increase of the particle contribution with the elasticity Capillary number (\ie deformability).

\section{Conclusion} \label{sec:conclusion}
We have implemented and validated a volume of fluid methodology based on the MTHINC method first proposed by \citet{ii_sugiyama_takeuchi_takagi_matsumoto_xiao_2012a} which can be used to study multiphase problems. A fully multidimensional hyperbolic tangent function is used to reconstruct the interface and the main advantages can be summarized as follows: \textit{i)} the geometric reconstruction of the interface is not required; \textit{ii)} both linear and quadratic surfaces can be easily constructed; \textit{iii)} the continuous multidimensional hyperbolic tangent function allows the direct calculations of the numerical fluxes, derivatives and normal vectors; \textit{iv)} the hyperbolic tangent function prevents the numerical diffusion that smears out the interface transition layer. Moreover, the implementation presented here allows the use of an efficient FFT-based fractional step method to solve the system of equations, and is based on the splitting of the pressure in a constant and a varying term, which makes the FFT solver applicable also when the density in the two phases differs.

We have studied the rheology of a system of two Newtonian fluids in a wall-bounded shear flow, \ie plane Couette flow, at low Reynolds number such that inertial effects are negligible. The rheology of the emulsion is analyzed by discussing how the effective viscosity $\mu$ is affected by variations of the volume fraction $\Phi$ and Capillary number $\Ca$. The effective viscosity $\mu$ is a non-linear function of both this parameters $\mu=\mu \left( \Phi, \Ca \right)$ and the emulsion shows an effective viscosity lower than the one of suspensions of rigid particles, deformable elastic particles and capsules. Differently form these other cases, here the droplets can merge, especially at high volume fractions and high Capillary numbers. Also, the effective viscosity curve has a negative second derivative with respect to $\Phi$, thus suggesting the presence of a maximum in the curve at higher volume fractions. The overall deformation of the emulsion has been quantified in terms of the interface surface area and the flow within the two phases studied by means of the flow topology parameter. In particular, we have shown that the flow of the suspending fluid is mainly a shear flow, while that of the dispersed fluid is more rotational.

Finally, we have analysed the contributions to the total shear stress of the two fluid phases and of the interfacial force and showed that the Reynolds stress contributions are negligible at this Reynolds number and that the viscous stress of the two fluids provide the dominant contribution. These have a non-uniform distribution across the channel, with the percentage contribution of both phases increasing with the volume fraction and Capillary number. An important contribution comes from the interfacial tension, whose effect increases with the volume fraction and decreases with the Capillary number.

The study presented in this work will be extended to consider different viscosity and density ratios between the two fluids, and to better understand the non-monotonic behavior of the effective viscosity and its effect on the rheological behavior of emulsions.


\begin{acknowledgements}
The work is supported by the Microflusa project. This effort receives funding from the European Union Horizon 2020 research and innovation program under Grant Agreement No. 664823. L.B. and M.E.R. also acknowledge financial support by the European Research Council grant, no. ERC-2013-CoG-616186, TRITOS. The computer time was provided by SNIC (Swedish National Infrastructure for Computing). 
\end{acknowledgements}

\bibliographystyle{spbasic}      
\bibliography{../../../../Articles/bibliography.bib}   

\end{document}

%% file: shape.tex
\begingroup
  \makeatletter
  \providecommand\color[2][]{%
    \GenericError{(gnuplot) \space\space\space\@spaces}{%
      Package color not loaded in conjunction with
      terminal option `colourtext'%
    }{See the gnuplot documentation for explanation.%
    }{Either use 'blacktext' in gnuplot or load the package
      color.sty in LaTeX.}%
    \renewcommand\color[2][]{}%
  }%
  \providecommand\includegraphics[2][]{%
    \GenericError{(gnuplot) \space\space\space\@spaces}{%
      Package graphicx or graphics not loaded%
    }{See the gnuplot documentation for explanation.%
    }{The gnuplot epslatex terminal needs graphicx.sty or graphics.sty.}%
    \renewcommand\includegraphics[2][]{}%
  }%
  \providecommand\rotatebox[2]{#2}%
  \@ifundefined{ifGPcolor}{%
    \newif\ifGPcolor
    \GPcolortrue
  }{}%
  \@ifundefined{ifGPblacktext}{%
    \newif\ifGPblacktext
    \GPblacktexttrue
  }{}%
  \let\gplgaddtomacro\g@addto@macro
  \gdef\gplbacktext{}%
  \gdef\gplfronttext{}%
  \makeatother
  \ifGPblacktext
    \def\colorrgb#1{}%
    \def\colorgray#1{}%
  \else
    \ifGPcolor
      \def\colorrgb#1{\color[rgb]{#1}}%
      \def\colorgray#1{\color[gray]{#1}}%
      \expandafter\def\csname LTw\endcsname{\color{white}}%
      \expandafter\def\csname LTb\endcsname{\color{black}}%
      \expandafter\def\csname LTa\endcsname{\color{black}}%
      \expandafter\def\csname LT0\endcsname{\color[rgb]{1,0,0}}%
      \expandafter\def\csname LT1\endcsname{\color[rgb]{0,1,0}}%
      \expandafter\def\csname LT2\endcsname{\color[rgb]{0,0,1}}%
      \expandafter\def\csname LT3\endcsname{\color[rgb]{1,0,1}}%
      \expandafter\def\csname LT4\endcsname{\color[rgb]{0,1,1}}%
      \expandafter\def\csname LT5\endcsname{\color[rgb]{1,1,0}}%
      \expandafter\def\csname LT6\endcsname{\color[rgb]{0,0,0}}%
      \expandafter\def\csname LT7\endcsname{\color[rgb]{1,0.3,0}}%
      \expandafter\def\csname LT8\endcsname{\color[rgb]{0.5,0.5,0.5}}%
    \else
      \def\colorrgb#1{\color{black}}%
      \def\colorgray#1{\color[gray]{#1}}%
      \expandafter\def\csname LTw\endcsname{\color{white}}%
      \expandafter\def\csname LTb\endcsname{\color{black}}%
      \expandafter\def\csname LTa\endcsname{\color{black}}%
      \expandafter\def\csname LT0\endcsname{\color{black}}%
      \expandafter\def\csname LT1\endcsname{\color{black}}%
      \expandafter\def\csname LT2\endcsname{\color{black}}%
      \expandafter\def\csname LT3\endcsname{\color{black}}%
      \expandafter\def\csname LT4\endcsname{\color{black}}%
      \expandafter\def\csname LT5\endcsname{\color{black}}%
      \expandafter\def\csname LT6\endcsname{\color{black}}%
      \expandafter\def\csname LT7\endcsname{\color{black}}%
      \expandafter\def\csname LT8\endcsname{\color{black}}%
    \fi
  \fi
    \setlength{\unitlength}{0.0500bp}%
    \ifx\gptboxheight\undefined%
      \newlength{\gptboxheight}%
      \newlength{\gptboxwidth}%
      \newsavebox{\gptboxtext}%
    \fi%
    \setlength{\fboxrule}{0.5pt}%
    \setlength{\fboxsep}{1pt}%
\begin{picture}(3310.00,2880.00)%
\definecolor{gpBackground}{rgb}{1.000, 1.000, 1.000}%
\put(0,0){\colorbox{gpBackground}{\makebox(3310.00,2880.00)[]{}}}%
    \gplgaddtomacro\gplbacktext{%
      \csname LTb\endcsname%
      \put(530,288){\makebox(0,0)[r]{\strut{}$-1$}}%
      \put(530,1555){\makebox(0,0)[r]{\strut{}$0$}}%
      \put(530,2821){\makebox(0,0)[r]{\strut{}$1$}}%
      \put(662,68){\makebox(0,0){\strut{}$-1$}}%
      \put(1903,68){\makebox(0,0){\strut{}$0$}}%
      \put(3143,68){\makebox(0,0){\strut{}$1$}}%
    }%
    \gplgaddtomacro\gplfronttext{%
      \csname LTb\endcsname%
      \put(24,1554){\rotatebox{-270}{\makebox(0,0){\strut{}$y/D$}}}%
      \put(1902,-262){\makebox(0,0){\strut{}$x/D$}}%
    }%
    \gplbacktext
    \put(0,0){\includegraphics{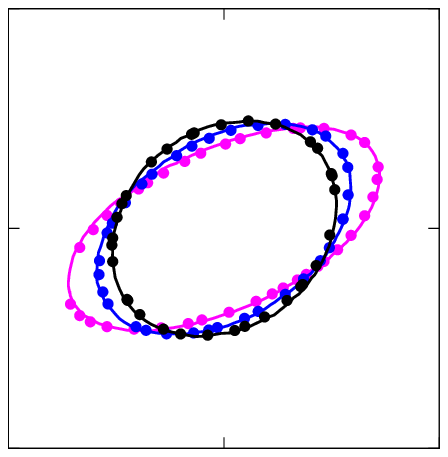}}%
    \gplfronttext
  \end{picture}%
\endgroup

%% file: bubble.tex
\begingroup
  \makeatletter
  \providecommand\color[2][]{%
    \GenericError{(gnuplot) \space\space\space\@spaces}{%
      Package color not loaded in conjunction with
      terminal option `colourtext'%
    }{See the gnuplot documentation for explanation.%
    }{Either use 'blacktext' in gnuplot or load the package
      color.sty in LaTeX.}%
    \renewcommand\color[2][]{}%
  }%
  \providecommand\includegraphics[2][]{%
    \GenericError{(gnuplot) \space\space\space\@spaces}{%
      Package graphicx or graphics not loaded%
    }{See the gnuplot documentation for explanation.%
    }{The gnuplot epslatex terminal needs graphicx.sty or graphics.sty.}%
    \renewcommand\includegraphics[2][]{}%
  }%
  \providecommand\rotatebox[2]{#2}%
  \@ifundefined{ifGPcolor}{%
    \newif\ifGPcolor
    \GPcolortrue
  }{}%
  \@ifundefined{ifGPblacktext}{%
    \newif\ifGPblacktext
    \GPblacktexttrue
  }{}%
  \let\gplgaddtomacro\g@addto@macro
  \gdef\gplbacktext{}%
  \gdef\gplfronttext{}%
  \makeatother
  \ifGPblacktext
    \def\colorrgb#1{}%
    \def\colorgray#1{}%
  \else
    \ifGPcolor
      \def\colorrgb#1{\color[rgb]{#1}}%
      \def\colorgray#1{\color[gray]{#1}}%
      \expandafter\def\csname LTw\endcsname{\color{white}}%
      \expandafter\def\csname LTb\endcsname{\color{black}}%
      \expandafter\def\csname LTa\endcsname{\color{black}}%
      \expandafter\def\csname LT0\endcsname{\color[rgb]{1,0,0}}%
      \expandafter\def\csname LT1\endcsname{\color[rgb]{0,1,0}}%
      \expandafter\def\csname LT2\endcsname{\color[rgb]{0,0,1}}%
      \expandafter\def\csname LT3\endcsname{\color[rgb]{1,0,1}}%
      \expandafter\def\csname LT4\endcsname{\color[rgb]{0,1,1}}%
      \expandafter\def\csname LT5\endcsname{\color[rgb]{1,1,0}}%
      \expandafter\def\csname LT6\endcsname{\color[rgb]{0,0,0}}%
      \expandafter\def\csname LT7\endcsname{\color[rgb]{1,0.3,0}}%
      \expandafter\def\csname LT8\endcsname{\color[rgb]{0.5,0.5,0.5}}%
    \else
      \def\colorrgb#1{\color{black}}%
      \def\colorgray#1{\color[gray]{#1}}%
      \expandafter\def\csname LTw\endcsname{\color{white}}%
      \expandafter\def\csname LTb\endcsname{\color{black}}%
      \expandafter\def\csname LTa\endcsname{\color{black}}%
      \expandafter\def\csname LT0\endcsname{\color{black}}%
      \expandafter\def\csname LT1\endcsname{\color{black}}%
      \expandafter\def\csname LT2\endcsname{\color{black}}%
      \expandafter\def\csname LT3\endcsname{\color{black}}%
      \expandafter\def\csname LT4\endcsname{\color{black}}%
      \expandafter\def\csname LT5\endcsname{\color{black}}%
      \expandafter\def\csname LT6\endcsname{\color{black}}%
      \expandafter\def\csname LT7\endcsname{\color{black}}%
      \expandafter\def\csname LT8\endcsname{\color{black}}%
    \fi
  \fi
    \setlength{\unitlength}{0.0500bp}%
    \ifx\gptboxheight\undefined%
      \newlength{\gptboxheight}%
      \newlength{\gptboxwidth}%
      \newsavebox{\gptboxtext}%
    \fi%
    \setlength{\fboxrule}{0.5pt}%
    \setlength{\fboxsep}{1pt}%
\begin{picture}(3310.00,2880.00)%
\definecolor{gpBackground}{rgb}{1.000, 1.000, 1.000}%
\put(0,0){\colorbox{gpBackground}{\makebox(3310.00,2880.00)[]{}}}%
    \gplgaddtomacro\gplbacktext{%
      \csname LTb\endcsname%
      \put(530,288){\makebox(0,0)[r]{\strut{}$0$}}%
      \put(530,2399){\makebox(0,0)[r]{\strut{}$1$}}%
      \put(662,68){\makebox(0,0){\strut{}$0$}}%
      \put(1654,68){\makebox(0,0){\strut{}$1$}}%
    }%
    \gplgaddtomacro\gplfronttext{%
      \csname LTb\endcsname%
      \put(376,1554){\rotatebox{-270}{\makebox(0,0){\strut{}}}}%
      \put(1902,-262){\makebox(0,0){\strut{}$t$}}%
    }%
    \gplbacktext
    \put(0,0){\includegraphics{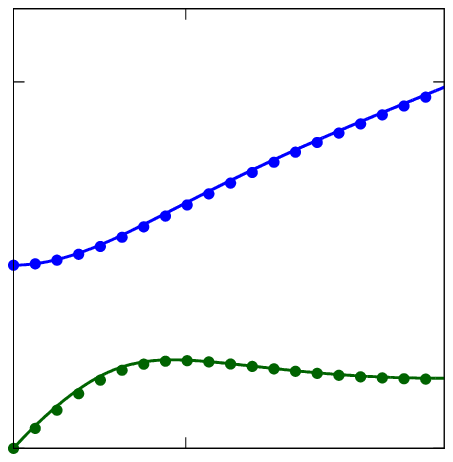}}%
    \gplfronttext
  \end{picture}%
\endgroup

%% file: vel.tex
\begingroup
  \makeatletter
  \providecommand\color[2][]{%
    \GenericError{(gnuplot) \space\space\space\@spaces}{%
      Package color not loaded in conjunction with
      terminal option `colourtext'%
    }{See the gnuplot documentation for explanation.%
    }{Either use 'blacktext' in gnuplot or load the package
      color.sty in LaTeX.}%
    \renewcommand\color[2][]{}%
  }%
  \providecommand\includegraphics[2][]{%
    \GenericError{(gnuplot) \space\space\space\@spaces}{%
      Package graphicx or graphics not loaded%
    }{See the gnuplot documentation for explanation.%
    }{The gnuplot epslatex terminal needs graphicx.sty or graphics.sty.}%
    \renewcommand\includegraphics[2][]{}%
  }%
  \providecommand\rotatebox[2]{#2}%
  \@ifundefined{ifGPcolor}{%
    \newif\ifGPcolor
    \GPcolortrue
  }{}%
  \@ifundefined{ifGPblacktext}{%
    \newif\ifGPblacktext
    \GPblacktexttrue
  }{}%
  \let\gplgaddtomacro\g@addto@macro
  \gdef\gplbacktext{}%
  \gdef\gplfronttext{}%
  \makeatother
  \ifGPblacktext
    \def\colorrgb#1{}%
    \def\colorgray#1{}%
  \else
    \ifGPcolor
      \def\colorrgb#1{\color[rgb]{#1}}%
      \def\colorgray#1{\color[gray]{#1}}%
      \expandafter\def\csname LTw\endcsname{\color{white}}%
      \expandafter\def\csname LTb\endcsname{\color{black}}%
      \expandafter\def\csname LTa\endcsname{\color{black}}%
      \expandafter\def\csname LT0\endcsname{\color[rgb]{1,0,0}}%
      \expandafter\def\csname LT1\endcsname{\color[rgb]{0,1,0}}%
      \expandafter\def\csname LT2\endcsname{\color[rgb]{0,0,1}}%
      \expandafter\def\csname LT3\endcsname{\color[rgb]{1,0,1}}%
      \expandafter\def\csname LT4\endcsname{\color[rgb]{0,1,1}}%
      \expandafter\def\csname LT5\endcsname{\color[rgb]{1,1,0}}%
      \expandafter\def\csname LT6\endcsname{\color[rgb]{0,0,0}}%
      \expandafter\def\csname LT7\endcsname{\color[rgb]{1,0.3,0}}%
      \expandafter\def\csname LT8\endcsname{\color[rgb]{0.5,0.5,0.5}}%
    \else
      \def\colorrgb#1{\color{black}}%
      \def\colorgray#1{\color[gray]{#1}}%
      \expandafter\def\csname LTw\endcsname{\color{white}}%
      \expandafter\def\csname LTb\endcsname{\color{black}}%
      \expandafter\def\csname LTa\endcsname{\color{black}}%
      \expandafter\def\csname LT0\endcsname{\color{black}}%
      \expandafter\def\csname LT1\endcsname{\color{black}}%
      \expandafter\def\csname LT2\endcsname{\color{black}}%
      \expandafter\def\csname LT3\endcsname{\color{black}}%
      \expandafter\def\csname LT4\endcsname{\color{black}}%
      \expandafter\def\csname LT5\endcsname{\color{black}}%
      \expandafter\def\csname LT6\endcsname{\color{black}}%
      \expandafter\def\csname LT7\endcsname{\color{black}}%
      \expandafter\def\csname LT8\endcsname{\color{black}}%
    \fi
  \fi
    \setlength{\unitlength}{0.0500bp}%
    \ifx\gptboxheight\undefined%
      \newlength{\gptboxheight}%
      \newlength{\gptboxwidth}%
      \newsavebox{\gptboxtext}%
    \fi%
    \setlength{\fboxrule}{0.5pt}%
    \setlength{\fboxsep}{1pt}%
\begin{picture}(3310.00,2880.00)%
\definecolor{gpBackground}{rgb}{1.000, 1.000, 1.000}%
\put(0,0){\colorbox{gpBackground}{\makebox(3310.00,2880.00)[]{}}}%
    \gplgaddtomacro\gplbacktext{%
      \csname LTb\endcsname%
      \put(530,288){\makebox(0,0)[r]{\strut{}$0$}}%
      \put(530,1555){\makebox(0,0)[r]{\strut{}$0.5$}}%
      \put(530,2821){\makebox(0,0)[r]{\strut{}$1$}}%
      \put(662,68){\makebox(0,0){\strut{}$0$}}%
      \put(1903,68){\makebox(0,0){\strut{}$0.5$}}%
      \put(3143,68){\makebox(0,0){\strut{}$1$}}%
    }%
    \gplgaddtomacro\gplfronttext{%
      \csname LTb\endcsname%
      \put(24,1554){\rotatebox{-270}{\makebox(0,0){\strut{}$\bra{\vf}$}}}%
      \put(1770,-262){\makebox(0,0){\strut{}$y/h$}}%
      \put(1902,2711){\makebox(0,0){\strut{}}}%
    }%
    \gplbacktext
    \put(0,0){\includegraphics{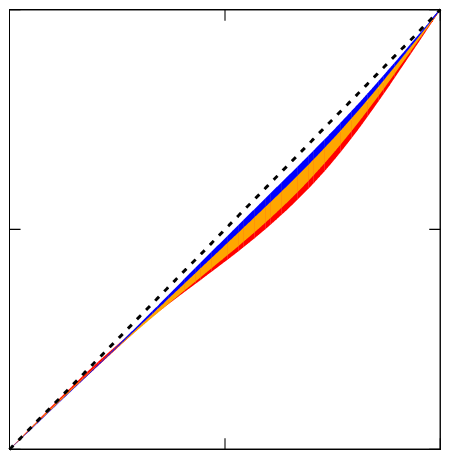}}%
    \gplfronttext
  \end{picture}%
\endgroup

%% file: vof.tex
\begingroup
  \makeatletter
  \providecommand\color[2][]{%
    \GenericError{(gnuplot) \space\space\space\@spaces}{%
      Package color not loaded in conjunction with
      terminal option `colourtext'%
    }{See the gnuplot documentation for explanation.%
    }{Either use 'blacktext' in gnuplot or load the package
      color.sty in LaTeX.}%
    \renewcommand\color[2][]{}%
  }%
  \providecommand\includegraphics[2][]{%
    \GenericError{(gnuplot) \space\space\space\@spaces}{%
      Package graphicx or graphics not loaded%
    }{See the gnuplot documentation for explanation.%
    }{The gnuplot epslatex terminal needs graphicx.sty or graphics.sty.}%
    \renewcommand\includegraphics[2][]{}%
  }%
  \providecommand\rotatebox[2]{#2}%
  \@ifundefined{ifGPcolor}{%
    \newif\ifGPcolor
    \GPcolortrue
  }{}%
  \@ifundefined{ifGPblacktext}{%
    \newif\ifGPblacktext
    \GPblacktexttrue
  }{}%
  \let\gplgaddtomacro\g@addto@macro
  \gdef\gplbacktext{}%
  \gdef\gplfronttext{}%
  \makeatother
  \ifGPblacktext
    \def\colorrgb#1{}%
    \def\colorgray#1{}%
  \else
    \ifGPcolor
      \def\colorrgb#1{\color[rgb]{#1}}%
      \def\colorgray#1{\color[gray]{#1}}%
      \expandafter\def\csname LTw\endcsname{\color{white}}%
      \expandafter\def\csname LTb\endcsname{\color{black}}%
      \expandafter\def\csname LTa\endcsname{\color{black}}%
      \expandafter\def\csname LT0\endcsname{\color[rgb]{1,0,0}}%
      \expandafter\def\csname LT1\endcsname{\color[rgb]{0,1,0}}%
      \expandafter\def\csname LT2\endcsname{\color[rgb]{0,0,1}}%
      \expandafter\def\csname LT3\endcsname{\color[rgb]{1,0,1}}%
      \expandafter\def\csname LT4\endcsname{\color[rgb]{0,1,1}}%
      \expandafter\def\csname LT5\endcsname{\color[rgb]{1,1,0}}%
      \expandafter\def\csname LT6\endcsname{\color[rgb]{0,0,0}}%
      \expandafter\def\csname LT7\endcsname{\color[rgb]{1,0.3,0}}%
      \expandafter\def\csname LT8\endcsname{\color[rgb]{0.5,0.5,0.5}}%
    \else
      \def\colorrgb#1{\color{black}}%
      \def\colorgray#1{\color[gray]{#1}}%
      \expandafter\def\csname LTw\endcsname{\color{white}}%
      \expandafter\def\csname LTb\endcsname{\color{black}}%
      \expandafter\def\csname LTa\endcsname{\color{black}}%
      \expandafter\def\csname LT0\endcsname{\color{black}}%
      \expandafter\def\csname LT1\endcsname{\color{black}}%
      \expandafter\def\csname LT2\endcsname{\color{black}}%
      \expandafter\def\csname LT3\endcsname{\color{black}}%
      \expandafter\def\csname LT4\endcsname{\color{black}}%
      \expandafter\def\csname LT5\endcsname{\color{black}}%
      \expandafter\def\csname LT6\endcsname{\color{black}}%
      \expandafter\def\csname LT7\endcsname{\color{black}}%
      \expandafter\def\csname LT8\endcsname{\color{black}}%
    \fi
  \fi
    \setlength{\unitlength}{0.0500bp}%
    \ifx\gptboxheight\undefined%
      \newlength{\gptboxheight}%
      \newlength{\gptboxwidth}%
      \newsavebox{\gptboxtext}%
    \fi%
    \setlength{\fboxrule}{0.5pt}%
    \setlength{\fboxsep}{1pt}%
\begin{picture}(3310.00,2880.00)%
\definecolor{gpBackground}{rgb}{1.000, 1.000, 1.000}%
\put(0,0){\colorbox{gpBackground}{\makebox(3310.00,2880.00)[]{}}}%
    \gplgaddtomacro\gplbacktext{%
      \csname LTb\endcsname%
      \put(530,288){\makebox(0,0)[r]{\strut{}$0$}}%
      \put(530,1132){\makebox(0,0)[r]{\strut{}$0.2$}}%
      \put(530,1977){\makebox(0,0)[r]{\strut{}$0.4$}}%
      \put(530,2821){\makebox(0,0)[r]{\strut{}$0.6$}}%
      \put(662,68){\makebox(0,0){\strut{}$0$}}%
      \put(1903,68){\makebox(0,0){\strut{}$0.5$}}%
      \put(3143,68){\makebox(0,0){\strut{}$1$}}%
    }%
    \gplgaddtomacro\gplfronttext{%
      \csname LTb\endcsname%
      \put(24,1554){\rotatebox{-270}{\makebox(0,0){\strut{}$\bra{\phis}$}}}%
      \put(2034,-262){\makebox(0,0){\strut{}$y/h$}}%
      \put(1902,2711){\makebox(0,0){\strut{}}}%
    }%
    \gplbacktext
    \put(0,0){\includegraphics{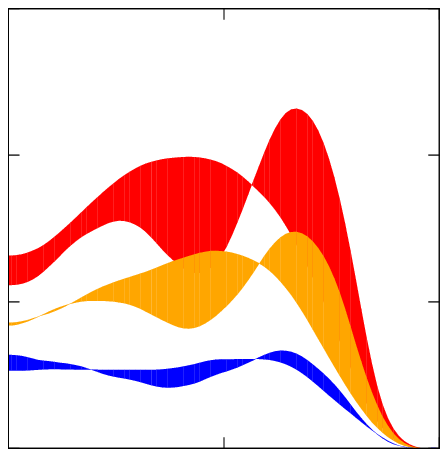}}%
    \gplfronttext
  \end{picture}%
\endgroup

%% file: viscF.tex
\begingroup
  \makeatletter
  \providecommand\color[2][]{%
    \GenericError{(gnuplot) \space\space\space\@spaces}{%
      Package color not loaded in conjunction with
      terminal option `colourtext'%
    }{See the gnuplot documentation for explanation.%
    }{Either use 'blacktext' in gnuplot or load the package
      color.sty in LaTeX.}%
    \renewcommand\color[2][]{}%
  }%
  \providecommand\includegraphics[2][]{%
    \GenericError{(gnuplot) \space\space\space\@spaces}{%
      Package graphicx or graphics not loaded%
    }{See the gnuplot documentation for explanation.%
    }{The gnuplot epslatex terminal needs graphicx.sty or graphics.sty.}%
    \renewcommand\includegraphics[2][]{}%
  }%
  \providecommand\rotatebox[2]{#2}%
  \@ifundefined{ifGPcolor}{%
    \newif\ifGPcolor
    \GPcolortrue
  }{}%
  \@ifundefined{ifGPblacktext}{%
    \newif\ifGPblacktext
    \GPblacktexttrue
  }{}%
  \let\gplgaddtomacro\g@addto@macro
  \gdef\gplbacktext{}%
  \gdef\gplfronttext{}%
  \makeatother
  \ifGPblacktext
    \def\colorrgb#1{}%
    \def\colorgray#1{}%
  \else
    \ifGPcolor
      \def\colorrgb#1{\color[rgb]{#1}}%
      \def\colorgray#1{\color[gray]{#1}}%
      \expandafter\def\csname LTw\endcsname{\color{white}}%
      \expandafter\def\csname LTb\endcsname{\color{black}}%
      \expandafter\def\csname LTa\endcsname{\color{black}}%
      \expandafter\def\csname LT0\endcsname{\color[rgb]{1,0,0}}%
      \expandafter\def\csname LT1\endcsname{\color[rgb]{0,1,0}}%
      \expandafter\def\csname LT2\endcsname{\color[rgb]{0,0,1}}%
      \expandafter\def\csname LT3\endcsname{\color[rgb]{1,0,1}}%
      \expandafter\def\csname LT4\endcsname{\color[rgb]{0,1,1}}%
      \expandafter\def\csname LT5\endcsname{\color[rgb]{1,1,0}}%
      \expandafter\def\csname LT6\endcsname{\color[rgb]{0,0,0}}%
      \expandafter\def\csname LT7\endcsname{\color[rgb]{1,0.3,0}}%
      \expandafter\def\csname LT8\endcsname{\color[rgb]{0.5,0.5,0.5}}%
    \else
      \def\colorrgb#1{\color{black}}%
      \def\colorgray#1{\color[gray]{#1}}%
      \expandafter\def\csname LTw\endcsname{\color{white}}%
      \expandafter\def\csname LTb\endcsname{\color{black}}%
      \expandafter\def\csname LTa\endcsname{\color{black}}%
      \expandafter\def\csname LT0\endcsname{\color{black}}%
      \expandafter\def\csname LT1\endcsname{\color{black}}%
      \expandafter\def\csname LT2\endcsname{\color{black}}%
      \expandafter\def\csname LT3\endcsname{\color{black}}%
      \expandafter\def\csname LT4\endcsname{\color{black}}%
      \expandafter\def\csname LT5\endcsname{\color{black}}%
      \expandafter\def\csname LT6\endcsname{\color{black}}%
      \expandafter\def\csname LT7\endcsname{\color{black}}%
      \expandafter\def\csname LT8\endcsname{\color{black}}%
    \fi
  \fi
    \setlength{\unitlength}{0.0500bp}%
    \ifx\gptboxheight\undefined%
      \newlength{\gptboxheight}%
      \newlength{\gptboxwidth}%
      \newsavebox{\gptboxtext}%
    \fi%
    \setlength{\fboxrule}{0.5pt}%
    \setlength{\fboxsep}{1pt}%
\begin{picture}(3310.00,2880.00)%
\definecolor{gpBackground}{rgb}{1.000, 1.000, 1.000}%
\put(0,0){\colorbox{gpBackground}{\makebox(3310.00,2880.00)[]{}}}%
    \gplgaddtomacro\gplbacktext{%
      \csname LTb\endcsname%
      \put(530,288){\makebox(0,0)[r]{\strut{}$1$}}%
      \put(530,1555){\makebox(0,0)[r]{\strut{}$1.2$}}%
      \put(530,2821){\makebox(0,0)[r]{\strut{}$1.4$}}%
      \put(662,68){\makebox(0,0){\strut{}$0$}}%
      \put(1371,68){\makebox(0,0){\strut{}$0.1$}}%
      \put(2080,68){\makebox(0,0){\strut{}$0.2$}}%
      \put(2789,68){\makebox(0,0){\strut{}$0.3$}}%
    }%
    \gplgaddtomacro\gplfronttext{%
      \csname LTb\endcsname%
      \put(24,1554){\rotatebox{-270}{\makebox(0,0){\strut{}$\mu/\muf$}}}%
      \put(1770,-262){\makebox(0,0){\strut{}$\Phi$}}%
      \put(1902,2711){\makebox(0,0){\strut{}}}%
    }%
    \gplbacktext
    \put(0,0){\includegraphics{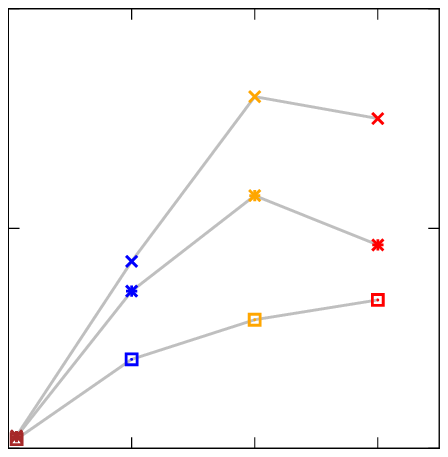}}%
    \gplfronttext
  \end{picture}%
\endgroup

%% file: viscT.tex
\begingroup
  \makeatletter
  \providecommand\color[2][]{%
    \GenericError{(gnuplot) \space\space\space\@spaces}{%
      Package color not loaded in conjunction with
      terminal option `colourtext'%
    }{See the gnuplot documentation for explanation.%
    }{Either use 'blacktext' in gnuplot or load the package
      color.sty in LaTeX.}%
    \renewcommand\color[2][]{}%
  }%
  \providecommand\includegraphics[2][]{%
    \GenericError{(gnuplot) \space\space\space\@spaces}{%
      Package graphicx or graphics not loaded%
    }{See the gnuplot documentation for explanation.%
    }{The gnuplot epslatex terminal needs graphicx.sty or graphics.sty.}%
    \renewcommand\includegraphics[2][]{}%
  }%
  \providecommand\rotatebox[2]{#2}%
  \@ifundefined{ifGPcolor}{%
    \newif\ifGPcolor
    \GPcolortrue
  }{}%
  \@ifundefined{ifGPblacktext}{%
    \newif\ifGPblacktext
    \GPblacktexttrue
  }{}%
  \let\gplgaddtomacro\g@addto@macro
  \gdef\gplbacktext{}%
  \gdef\gplfronttext{}%
  \makeatother
  \ifGPblacktext
    \def\colorrgb#1{}%
    \def\colorgray#1{}%
  \else
    \ifGPcolor
      \def\colorrgb#1{\color[rgb]{#1}}%
      \def\colorgray#1{\color[gray]{#1}}%
      \expandafter\def\csname LTw\endcsname{\color{white}}%
      \expandafter\def\csname LTb\endcsname{\color{black}}%
      \expandafter\def\csname LTa\endcsname{\color{black}}%
      \expandafter\def\csname LT0\endcsname{\color[rgb]{1,0,0}}%
      \expandafter\def\csname LT1\endcsname{\color[rgb]{0,1,0}}%
      \expandafter\def\csname LT2\endcsname{\color[rgb]{0,0,1}}%
      \expandafter\def\csname LT3\endcsname{\color[rgb]{1,0,1}}%
      \expandafter\def\csname LT4\endcsname{\color[rgb]{0,1,1}}%
      \expandafter\def\csname LT5\endcsname{\color[rgb]{1,1,0}}%
      \expandafter\def\csname LT6\endcsname{\color[rgb]{0,0,0}}%
      \expandafter\def\csname LT7\endcsname{\color[rgb]{1,0.3,0}}%
      \expandafter\def\csname LT8\endcsname{\color[rgb]{0.5,0.5,0.5}}%
    \else
      \def\colorrgb#1{\color{black}}%
      \def\colorgray#1{\color[gray]{#1}}%
      \expandafter\def\csname LTw\endcsname{\color{white}}%
      \expandafter\def\csname LTb\endcsname{\color{black}}%
      \expandafter\def\csname LTa\endcsname{\color{black}}%
      \expandafter\def\csname LT0\endcsname{\color{black}}%
      \expandafter\def\csname LT1\endcsname{\color{black}}%
      \expandafter\def\csname LT2\endcsname{\color{black}}%
      \expandafter\def\csname LT3\endcsname{\color{black}}%
      \expandafter\def\csname LT4\endcsname{\color{black}}%
      \expandafter\def\csname LT5\endcsname{\color{black}}%
      \expandafter\def\csname LT6\endcsname{\color{black}}%
      \expandafter\def\csname LT7\endcsname{\color{black}}%
      \expandafter\def\csname LT8\endcsname{\color{black}}%
    \fi
  \fi
    \setlength{\unitlength}{0.0500bp}%
    \ifx\gptboxheight\undefined%
      \newlength{\gptboxheight}%
      \newlength{\gptboxwidth}%
      \newsavebox{\gptboxtext}%
    \fi%
    \setlength{\fboxrule}{0.5pt}%
    \setlength{\fboxsep}{1pt}%
\begin{picture}(3310.00,2880.00)%
\definecolor{gpBackground}{rgb}{1.000, 1.000, 1.000}%
\put(0,0){\colorbox{gpBackground}{\makebox(3310.00,2880.00)[]{}}}%
    \gplgaddtomacro\gplbacktext{%
      \csname LTb\endcsname%
      \put(530,288){\makebox(0,0)[r]{\strut{}$1$}}%
      \put(530,1555){\makebox(0,0)[r]{\strut{}$2$}}%
      \put(530,2821){\makebox(0,0)[r]{\strut{}$3$}}%
      \put(662,68){\makebox(0,0){\strut{}$0$}}%
      \put(1696,68){\makebox(0,0){\strut{}$50$}}%
      \put(2730,68){\makebox(0,0){\strut{}$100$}}%
    }%
    \gplgaddtomacro\gplfronttext{%
      \csname LTb\endcsname%
      \put(288,1554){\rotatebox{-270}{\makebox(0,0){\strut{}$\mu/\muf$}}}%
      \put(1770,-262){\makebox(0,0){\strut{}$\dot{\gamma}t$}}%
      \put(1902,2711){\makebox(0,0){\strut{}}}%
    }%
    \gplbacktext
    \put(0,0){\includegraphics{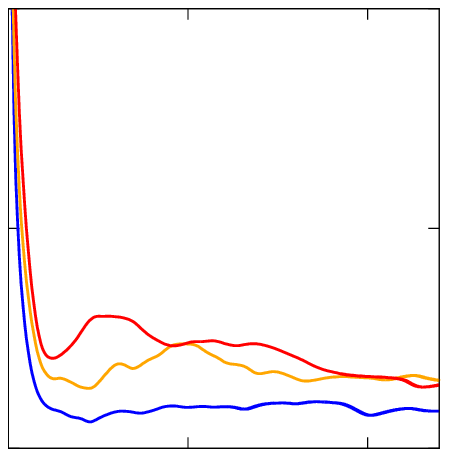}}%
    \gplfronttext
  \end{picture}%
\endgroup

%% file: viscALL.tex
\begingroup
  \makeatletter
  \providecommand\color[2][]{%
    \GenericError{(gnuplot) \space\space\space\@spaces}{%
      Package color not loaded in conjunction with
      terminal option `colourtext'%
    }{See the gnuplot documentation for explanation.%
    }{Either use 'blacktext' in gnuplot or load the package
      color.sty in LaTeX.}%
    \renewcommand\color[2][]{}%
  }%
  \providecommand\includegraphics[2][]{%
    \GenericError{(gnuplot) \space\space\space\@spaces}{%
      Package graphicx or graphics not loaded%
    }{See the gnuplot documentation for explanation.%
    }{The gnuplot epslatex terminal needs graphicx.sty or graphics.sty.}%
    \renewcommand\includegraphics[2][]{}%
  }%
  \providecommand\rotatebox[2]{#2}%
  \@ifundefined{ifGPcolor}{%
    \newif\ifGPcolor
    \GPcolortrue
  }{}%
  \@ifundefined{ifGPblacktext}{%
    \newif\ifGPblacktext
    \GPblacktexttrue
  }{}%
  \let\gplgaddtomacro\g@addto@macro
  \gdef\gplbacktext{}%
  \gdef\gplfronttext{}%
  \makeatother
  \ifGPblacktext
    \def\colorrgb#1{}%
    \def\colorgray#1{}%
  \else
    \ifGPcolor
      \def\colorrgb#1{\color[rgb]{#1}}%
      \def\colorgray#1{\color[gray]{#1}}%
      \expandafter\def\csname LTw\endcsname{\color{white}}%
      \expandafter\def\csname LTb\endcsname{\color{black}}%
      \expandafter\def\csname LTa\endcsname{\color{black}}%
      \expandafter\def\csname LT0\endcsname{\color[rgb]{1,0,0}}%
      \expandafter\def\csname LT1\endcsname{\color[rgb]{0,1,0}}%
      \expandafter\def\csname LT2\endcsname{\color[rgb]{0,0,1}}%
      \expandafter\def\csname LT3\endcsname{\color[rgb]{1,0,1}}%
      \expandafter\def\csname LT4\endcsname{\color[rgb]{0,1,1}}%
      \expandafter\def\csname LT5\endcsname{\color[rgb]{1,1,0}}%
      \expandafter\def\csname LT6\endcsname{\color[rgb]{0,0,0}}%
      \expandafter\def\csname LT7\endcsname{\color[rgb]{1,0.3,0}}%
      \expandafter\def\csname LT8\endcsname{\color[rgb]{0.5,0.5,0.5}}%
    \else
      \def\colorrgb#1{\color{black}}%
      \def\colorgray#1{\color[gray]{#1}}%
      \expandafter\def\csname LTw\endcsname{\color{white}}%
      \expandafter\def\csname LTb\endcsname{\color{black}}%
      \expandafter\def\csname LTa\endcsname{\color{black}}%
      \expandafter\def\csname LT0\endcsname{\color{black}}%
      \expandafter\def\csname LT1\endcsname{\color{black}}%
      \expandafter\def\csname LT2\endcsname{\color{black}}%
      \expandafter\def\csname LT3\endcsname{\color{black}}%
      \expandafter\def\csname LT4\endcsname{\color{black}}%
      \expandafter\def\csname LT5\endcsname{\color{black}}%
      \expandafter\def\csname LT6\endcsname{\color{black}}%
      \expandafter\def\csname LT7\endcsname{\color{black}}%
      \expandafter\def\csname LT8\endcsname{\color{black}}%
    \fi
  \fi
    \setlength{\unitlength}{0.0500bp}%
    \ifx\gptboxheight\undefined%
      \newlength{\gptboxheight}%
      \newlength{\gptboxwidth}%
      \newsavebox{\gptboxtext}%
    \fi%
    \setlength{\fboxrule}{0.5pt}%
    \setlength{\fboxsep}{1pt}%
\begin{picture}(6622.00,2880.00)%
\definecolor{gpBackground}{rgb}{1.000, 1.000, 1.000}%
\put(0,0){\colorbox{gpBackground}{\makebox(6622.00,2880.00)[]{}}}%
    \gplgaddtomacro\gplbacktext{%
      \csname LTb\endcsname%
      \put(530,288){\makebox(0,0)[r]{\strut{}$1$}}%
      \put(530,1132){\makebox(0,0)[r]{\strut{}$2$}}%
      \put(530,1977){\makebox(0,0)[r]{\strut{}$3$}}%
      \put(530,2821){\makebox(0,0)[r]{\strut{}$4$}}%
      \put(662,68){\makebox(0,0){\strut{}$0$}}%
      \put(2175,68){\makebox(0,0){\strut{}$0.1$}}%
      \put(3688,68){\makebox(0,0){\strut{}$0.2$}}%
      \put(5201,68){\makebox(0,0){\strut{}$0.3$}}%
    }%
    \gplgaddtomacro\gplfronttext{%
      \csname LTb\endcsname%
      \put(288,1554){\rotatebox{-270}{\makebox(0,0){\strut{}$\mu/\muf$}}}%
      \put(3178,-262){\makebox(0,0){\strut{}$\Phi$}}%
      \put(3310,2711){\makebox(0,0){\strut{}}}%
    }%
    \gplbacktext
    \put(0,0){\includegraphics{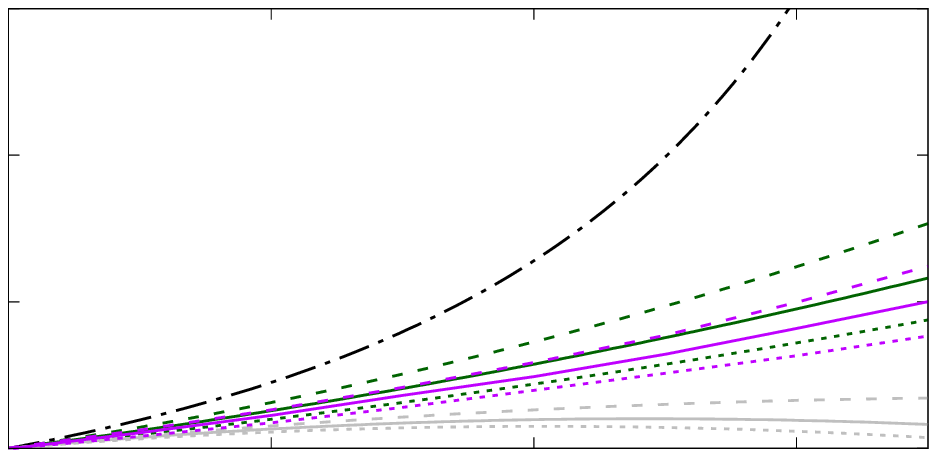}}%
    \gplfronttext
  \end{picture}%
\endgroup

%% file: SF.tex
\begingroup
  \makeatletter
  \providecommand\color[2][]{%
    \GenericError{(gnuplot) \space\space\space\@spaces}{%
      Package color not loaded in conjunction with
      terminal option `colourtext'%
    }{See the gnuplot documentation for explanation.%
    }{Either use 'blacktext' in gnuplot or load the package
      color.sty in LaTeX.}%
    \renewcommand\color[2][]{}%
  }%
  \providecommand\includegraphics[2][]{%
    \GenericError{(gnuplot) \space\space\space\@spaces}{%
      Package graphicx or graphics not loaded%
    }{See the gnuplot documentation for explanation.%
    }{The gnuplot epslatex terminal needs graphicx.sty or graphics.sty.}%
    \renewcommand\includegraphics[2][]{}%
  }%
  \providecommand\rotatebox[2]{#2}%
  \@ifundefined{ifGPcolor}{%
    \newif\ifGPcolor
    \GPcolortrue
  }{}%
  \@ifundefined{ifGPblacktext}{%
    \newif\ifGPblacktext
    \GPblacktexttrue
  }{}%
  \let\gplgaddtomacro\g@addto@macro
  \gdef\gplbacktext{}%
  \gdef\gplfronttext{}%
  \makeatother
  \ifGPblacktext
    \def\colorrgb#1{}%
    \def\colorgray#1{}%
  \else
    \ifGPcolor
      \def\colorrgb#1{\color[rgb]{#1}}%
      \def\colorgray#1{\color[gray]{#1}}%
      \expandafter\def\csname LTw\endcsname{\color{white}}%
      \expandafter\def\csname LTb\endcsname{\color{black}}%
      \expandafter\def\csname LTa\endcsname{\color{black}}%
      \expandafter\def\csname LT0\endcsname{\color[rgb]{1,0,0}}%
      \expandafter\def\csname LT1\endcsname{\color[rgb]{0,1,0}}%
      \expandafter\def\csname LT2\endcsname{\color[rgb]{0,0,1}}%
      \expandafter\def\csname LT3\endcsname{\color[rgb]{1,0,1}}%
      \expandafter\def\csname LT4\endcsname{\color[rgb]{0,1,1}}%
      \expandafter\def\csname LT5\endcsname{\color[rgb]{1,1,0}}%
      \expandafter\def\csname LT6\endcsname{\color[rgb]{0,0,0}}%
      \expandafter\def\csname LT7\endcsname{\color[rgb]{1,0.3,0}}%
      \expandafter\def\csname LT8\endcsname{\color[rgb]{0.5,0.5,0.5}}%
    \else
      \def\colorrgb#1{\color{black}}%
      \def\colorgray#1{\color[gray]{#1}}%
      \expandafter\def\csname LTw\endcsname{\color{white}}%
      \expandafter\def\csname LTb\endcsname{\color{black}}%
      \expandafter\def\csname LTa\endcsname{\color{black}}%
      \expandafter\def\csname LT0\endcsname{\color{black}}%
      \expandafter\def\csname LT1\endcsname{\color{black}}%
      \expandafter\def\csname LT2\endcsname{\color{black}}%
      \expandafter\def\csname LT3\endcsname{\color{black}}%
      \expandafter\def\csname LT4\endcsname{\color{black}}%
      \expandafter\def\csname LT5\endcsname{\color{black}}%
      \expandafter\def\csname LT6\endcsname{\color{black}}%
      \expandafter\def\csname LT7\endcsname{\color{black}}%
      \expandafter\def\csname LT8\endcsname{\color{black}}%
    \fi
  \fi
    \setlength{\unitlength}{0.0500bp}%
    \ifx\gptboxheight\undefined%
      \newlength{\gptboxheight}%
      \newlength{\gptboxwidth}%
      \newsavebox{\gptboxtext}%
    \fi%
    \setlength{\fboxrule}{0.5pt}%
    \setlength{\fboxsep}{1pt}%
\begin{picture}(3310.00,2880.00)%
\definecolor{gpBackground}{rgb}{1.000, 1.000, 1.000}%
\put(0,0){\colorbox{gpBackground}{\makebox(3310.00,2880.00)[]{}}}%
    \gplgaddtomacro\gplbacktext{%
      \csname LTb\endcsname%
      \put(530,288){\makebox(0,0)[r]{\strut{}$0.3$}}%
      \put(530,1132){\makebox(0,0)[r]{\strut{}$0.6$}}%
      \put(530,1977){\makebox(0,0)[r]{\strut{}$0.9$}}%
      \put(530,2821){\makebox(0,0)[r]{\strut{}$1.2$}}%
      \put(662,68){\makebox(0,0){\strut{}$0$}}%
      \put(1371,68){\makebox(0,0){\strut{}$0.1$}}%
      \put(2080,68){\makebox(0,0){\strut{}$0.2$}}%
      \put(2789,68){\makebox(0,0){\strut{}$0.3$}}%
    }%
    \gplgaddtomacro\gplfronttext{%
      \csname LTb\endcsname%
      \put(24,1554){\rotatebox{-270}{\makebox(0,0){\strut{}$\mathcal{S}/\mathcal{S}_0$}}}%
      \put(1770,-262){\makebox(0,0){\strut{}$\Phi$}}%
      \put(1902,2711){\makebox(0,0){\strut{}}}%
    }%
    \gplbacktext
    \put(0,0){\includegraphics{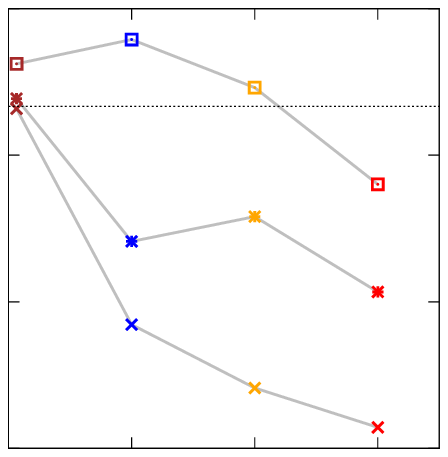}}%
    \gplfronttext
  \end{picture}%
\endgroup

%% file: pdf.tex
\begingroup
  \makeatletter
  \providecommand\color[2][]{%
    \GenericError{(gnuplot) \space\space\space\@spaces}{%
      Package color not loaded in conjunction with
      terminal option `colourtext'%
    }{See the gnuplot documentation for explanation.%
    }{Either use 'blacktext' in gnuplot or load the package
      color.sty in LaTeX.}%
    \renewcommand\color[2][]{}%
  }%
  \providecommand\includegraphics[2][]{%
    \GenericError{(gnuplot) \space\space\space\@spaces}{%
      Package graphicx or graphics not loaded%
    }{See the gnuplot documentation for explanation.%
    }{The gnuplot epslatex terminal needs graphicx.sty or graphics.sty.}%
    \renewcommand\includegraphics[2][]{}%
  }%
  \providecommand\rotatebox[2]{#2}%
  \@ifundefined{ifGPcolor}{%
    \newif\ifGPcolor
    \GPcolortrue
  }{}%
  \@ifundefined{ifGPblacktext}{%
    \newif\ifGPblacktext
    \GPblacktexttrue
  }{}%
  \let\gplgaddtomacro\g@addto@macro
  \gdef\gplbacktext{}%
  \gdef\gplfronttext{}%
  \makeatother
  \ifGPblacktext
    \def\colorrgb#1{}%
    \def\colorgray#1{}%
  \else
    \ifGPcolor
      \def\colorrgb#1{\color[rgb]{#1}}%
      \def\colorgray#1{\color[gray]{#1}}%
      \expandafter\def\csname LTw\endcsname{\color{white}}%
      \expandafter\def\csname LTb\endcsname{\color{black}}%
      \expandafter\def\csname LTa\endcsname{\color{black}}%
      \expandafter\def\csname LT0\endcsname{\color[rgb]{1,0,0}}%
      \expandafter\def\csname LT1\endcsname{\color[rgb]{0,1,0}}%
      \expandafter\def\csname LT2\endcsname{\color[rgb]{0,0,1}}%
      \expandafter\def\csname LT3\endcsname{\color[rgb]{1,0,1}}%
      \expandafter\def\csname LT4\endcsname{\color[rgb]{0,1,1}}%
      \expandafter\def\csname LT5\endcsname{\color[rgb]{1,1,0}}%
      \expandafter\def\csname LT6\endcsname{\color[rgb]{0,0,0}}%
      \expandafter\def\csname LT7\endcsname{\color[rgb]{1,0.3,0}}%
      \expandafter\def\csname LT8\endcsname{\color[rgb]{0.5,0.5,0.5}}%
    \else
      \def\colorrgb#1{\color{black}}%
      \def\colorgray#1{\color[gray]{#1}}%
      \expandafter\def\csname LTw\endcsname{\color{white}}%
      \expandafter\def\csname LTb\endcsname{\color{black}}%
      \expandafter\def\csname LTa\endcsname{\color{black}}%
      \expandafter\def\csname LT0\endcsname{\color{black}}%
      \expandafter\def\csname LT1\endcsname{\color{black}}%
      \expandafter\def\csname LT2\endcsname{\color{black}}%
      \expandafter\def\csname LT3\endcsname{\color{black}}%
      \expandafter\def\csname LT4\endcsname{\color{black}}%
      \expandafter\def\csname LT5\endcsname{\color{black}}%
      \expandafter\def\csname LT6\endcsname{\color{black}}%
      \expandafter\def\csname LT7\endcsname{\color{black}}%
      \expandafter\def\csname LT8\endcsname{\color{black}}%
    \fi
  \fi
    \setlength{\unitlength}{0.0500bp}%
    \ifx\gptboxheight\undefined%
      \newlength{\gptboxheight}%
      \newlength{\gptboxwidth}%
      \newsavebox{\gptboxtext}%
    \fi%
    \setlength{\fboxrule}{0.5pt}%
    \setlength{\fboxsep}{1pt}%
\begin{picture}(3310.00,2880.00)%
\definecolor{gpBackground}{rgb}{1.000, 1.000, 1.000}%
\put(0,0){\colorbox{gpBackground}{\makebox(3310.00,2880.00)[]{}}}%
    \gplgaddtomacro\gplbacktext{%
      \csname LTb\endcsname%
      \put(530,288){\makebox(0,0)[r]{\strut{}$0$}}%
      \put(530,1555){\makebox(0,0)[r]{\strut{}$0.1$}}%
      \put(530,2821){\makebox(0,0)[r]{\strut{}$0.2$}}%
      \put(662,68){\makebox(0,0){\strut{}$-1$}}%
      \put(1282,68){\makebox(0,0){\strut{}$-0.5$}}%
      \put(1903,68){\makebox(0,0){\strut{}$0$}}%
      \put(2523,68){\makebox(0,0){\strut{}$0.5$}}%
      \put(3143,68){\makebox(0,0){\strut{}$1$}}%
    }%
    \gplgaddtomacro\gplfronttext{%
      \csname LTb\endcsname%
      \put(24,1554){\rotatebox{-270}{\makebox(0,0){\strut{}$pdf$}}}%
      \put(1770,-262){\makebox(0,0){\strut{}$\mathcal{Q}$}}%
      \put(1902,2711){\makebox(0,0){\strut{}}}%
    }%
    \gplbacktext
    \put(0,0){\includegraphics{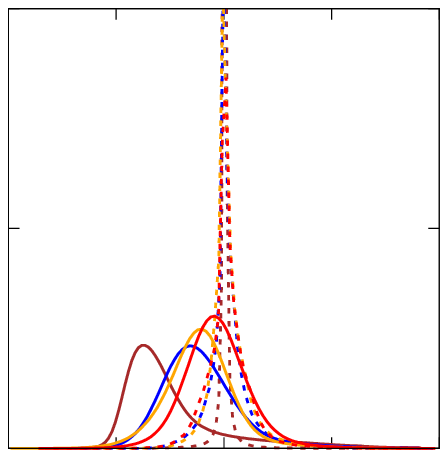}}%
    \gplfronttext
  \end{picture}%
\endgroup

%% file: tauG11.tex
\begingroup
  \makeatletter
  \providecommand\color[2][]{%
    \GenericError{(gnuplot) \space\space\space\@spaces}{%
      Package color not loaded in conjunction with
      terminal option `colourtext'%
    }{See the gnuplot documentation for explanation.%
    }{Either use 'blacktext' in gnuplot or load the package
      color.sty in LaTeX.}%
    \renewcommand\color[2][]{}%
  }%
  \providecommand\includegraphics[2][]{%
    \GenericError{(gnuplot) \space\space\space\@spaces}{%
      Package graphicx or graphics not loaded%
    }{See the gnuplot documentation for explanation.%
    }{The gnuplot epslatex terminal needs graphicx.sty or graphics.sty.}%
    \renewcommand\includegraphics[2][]{}%
  }%
  \providecommand\rotatebox[2]{#2}%
  \@ifundefined{ifGPcolor}{%
    \newif\ifGPcolor
    \GPcolortrue
  }{}%
  \@ifundefined{ifGPblacktext}{%
    \newif\ifGPblacktext
    \GPblacktexttrue
  }{}%
  \let\gplgaddtomacro\g@addto@macro
  \gdef\gplbacktext{}%
  \gdef\gplfronttext{}%
  \makeatother
  \ifGPblacktext
    \def\colorrgb#1{}%
    \def\colorgray#1{}%
  \else
    \ifGPcolor
      \def\colorrgb#1{\color[rgb]{#1}}%
      \def\colorgray#1{\color[gray]{#1}}%
      \expandafter\def\csname LTw\endcsname{\color{white}}%
      \expandafter\def\csname LTb\endcsname{\color{black}}%
      \expandafter\def\csname LTa\endcsname{\color{black}}%
      \expandafter\def\csname LT0\endcsname{\color[rgb]{1,0,0}}%
      \expandafter\def\csname LT1\endcsname{\color[rgb]{0,1,0}}%
      \expandafter\def\csname LT2\endcsname{\color[rgb]{0,0,1}}%
      \expandafter\def\csname LT3\endcsname{\color[rgb]{1,0,1}}%
      \expandafter\def\csname LT4\endcsname{\color[rgb]{0,1,1}}%
      \expandafter\def\csname LT5\endcsname{\color[rgb]{1,1,0}}%
      \expandafter\def\csname LT6\endcsname{\color[rgb]{0,0,0}}%
      \expandafter\def\csname LT7\endcsname{\color[rgb]{1,0.3,0}}%
      \expandafter\def\csname LT8\endcsname{\color[rgb]{0.5,0.5,0.5}}%
    \else
      \def\colorrgb#1{\color{black}}%
      \def\colorgray#1{\color[gray]{#1}}%
      \expandafter\def\csname LTw\endcsname{\color{white}}%
      \expandafter\def\csname LTb\endcsname{\color{black}}%
      \expandafter\def\csname LTa\endcsname{\color{black}}%
      \expandafter\def\csname LT0\endcsname{\color{black}}%
      \expandafter\def\csname LT1\endcsname{\color{black}}%
      \expandafter\def\csname LT2\endcsname{\color{black}}%
      \expandafter\def\csname LT3\endcsname{\color{black}}%
      \expandafter\def\csname LT4\endcsname{\color{black}}%
      \expandafter\def\csname LT5\endcsname{\color{black}}%
      \expandafter\def\csname LT6\endcsname{\color{black}}%
      \expandafter\def\csname LT7\endcsname{\color{black}}%
      \expandafter\def\csname LT8\endcsname{\color{black}}%
    \fi
  \fi
    \setlength{\unitlength}{0.0500bp}%
    \ifx\gptboxheight\undefined%
      \newlength{\gptboxheight}%
      \newlength{\gptboxwidth}%
      \newsavebox{\gptboxtext}%
    \fi%
    \setlength{\fboxrule}{0.5pt}%
    \setlength{\fboxsep}{1pt}%
\begin{picture}(3310.00,2880.00)%
\definecolor{gpBackground}{rgb}{1.000, 1.000, 1.000}%
\put(0,0){\colorbox{gpBackground}{\makebox(3310.00,2880.00)[]{}}}%
    \gplgaddtomacro\gplbacktext{%
      \csname LTb\endcsname%
      \put(530,288){\makebox(0,0)[r]{\strut{}$0$}}%
      \put(530,1555){\makebox(0,0)[r]{\strut{}$0.5$}}%
      \put(530,2821){\makebox(0,0)[r]{\strut{}$1$}}%
      \put(662,68){\makebox(0,0){\strut{}$0$}}%
      \put(1903,68){\makebox(0,0){\strut{}$0.5$}}%
      \put(3143,68){\makebox(0,0){\strut{}$1$}}%
    }%
    \gplgaddtomacro\gplfronttext{%
      \csname LTb\endcsname%
      \put(24,1554){\rotatebox{-270}{\makebox(0,0){\strut{}$\bra{\sigma_{12}}$}}}%
      \put(1770,-262){\makebox(0,0){\strut{}$y/h$}}%
      \put(1902,2711){\makebox(0,0){\strut{}}}%
    }%
    \gplbacktext
    \put(0,0){\includegraphics{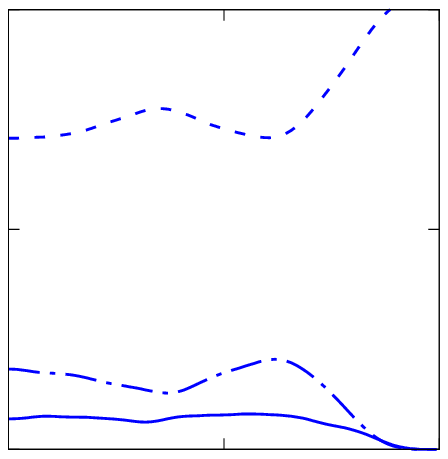}}%
    \gplfronttext
  \end{picture}%
\endgroup

%% file: tauhist1.tex
\begingroup
  \makeatletter
  \providecommand\color[2][]{%
    \GenericError{(gnuplot) \space\space\space\@spaces}{%
      Package color not loaded in conjunction with
      terminal option `colourtext'%
    }{See the gnuplot documentation for explanation.%
    }{Either use 'blacktext' in gnuplot or load the package
      color.sty in LaTeX.}%
    \renewcommand\color[2][]{}%
  }%
  \providecommand\includegraphics[2][]{%
    \GenericError{(gnuplot) \space\space\space\@spaces}{%
      Package graphicx or graphics not loaded%
    }{See the gnuplot documentation for explanation.%
    }{The gnuplot epslatex terminal needs graphicx.sty or graphics.sty.}%
    \renewcommand\includegraphics[2][]{}%
  }%
  \providecommand\rotatebox[2]{#2}%
  \@ifundefined{ifGPcolor}{%
    \newif\ifGPcolor
    \GPcolortrue
  }{}%
  \@ifundefined{ifGPblacktext}{%
    \newif\ifGPblacktext
    \GPblacktexttrue
  }{}%
  \let\gplgaddtomacro\g@addto@macro
  \gdef\gplbacktext{}%
  \gdef\gplfronttext{}%
  \makeatother
  \ifGPblacktext
    \def\colorrgb#1{}%
    \def\colorgray#1{}%
  \else
    \ifGPcolor
      \def\colorrgb#1{\color[rgb]{#1}}%
      \def\colorgray#1{\color[gray]{#1}}%
      \expandafter\def\csname LTw\endcsname{\color{white}}%
      \expandafter\def\csname LTb\endcsname{\color{black}}%
      \expandafter\def\csname LTa\endcsname{\color{black}}%
      \expandafter\def\csname LT0\endcsname{\color[rgb]{1,0,0}}%
      \expandafter\def\csname LT1\endcsname{\color[rgb]{0,1,0}}%
      \expandafter\def\csname LT2\endcsname{\color[rgb]{0,0,1}}%
      \expandafter\def\csname LT3\endcsname{\color[rgb]{1,0,1}}%
      \expandafter\def\csname LT4\endcsname{\color[rgb]{0,1,1}}%
      \expandafter\def\csname LT5\endcsname{\color[rgb]{1,1,0}}%
      \expandafter\def\csname LT6\endcsname{\color[rgb]{0,0,0}}%
      \expandafter\def\csname LT7\endcsname{\color[rgb]{1,0.3,0}}%
      \expandafter\def\csname LT8\endcsname{\color[rgb]{0.5,0.5,0.5}}%
    \else
      \def\colorrgb#1{\color{black}}%
      \def\colorgray#1{\color[gray]{#1}}%
      \expandafter\def\csname LTw\endcsname{\color{white}}%
      \expandafter\def\csname LTb\endcsname{\color{black}}%
      \expandafter\def\csname LTa\endcsname{\color{black}}%
      \expandafter\def\csname LT0\endcsname{\color{black}}%
      \expandafter\def\csname LT1\endcsname{\color{black}}%
      \expandafter\def\csname LT2\endcsname{\color{black}}%
      \expandafter\def\csname LT3\endcsname{\color{black}}%
      \expandafter\def\csname LT4\endcsname{\color{black}}%
      \expandafter\def\csname LT5\endcsname{\color{black}}%
      \expandafter\def\csname LT6\endcsname{\color{black}}%
      \expandafter\def\csname LT7\endcsname{\color{black}}%
      \expandafter\def\csname LT8\endcsname{\color{black}}%
    \fi
  \fi
    \setlength{\unitlength}{0.0500bp}%
    \ifx\gptboxheight\undefined%
      \newlength{\gptboxheight}%
      \newlength{\gptboxwidth}%
      \newsavebox{\gptboxtext}%
    \fi%
    \setlength{\fboxrule}{0.5pt}%
    \setlength{\fboxsep}{1pt}%
\begin{picture}(3310.00,2880.00)%
\definecolor{gpBackground}{rgb}{1.000, 1.000, 1.000}%
\put(0,0){\colorbox{gpBackground}{\makebox(3310.00,2880.00)[]{}}}%
    \gplgaddtomacro\gplbacktext{%
      \csname LTb\endcsname%
      \put(530,288){\makebox(0,0)[r]{\strut{}$0$}}%
      \put(530,921){\makebox(0,0)[r]{\strut{}$25$}}%
      \put(530,1555){\makebox(0,0)[r]{\strut{}$50$}}%
      \put(530,2188){\makebox(0,0)[r]{\strut{}$75$}}%
      \put(530,2821){\makebox(0,0)[r]{\strut{}$100$}}%
      \put(1076,68){\makebox(0,0){\strut{}0.1}}%
      \put(1903,68){\makebox(0,0){\strut{}0.2}}%
      \put(2730,68){\makebox(0,0){\strut{}0.4}}%
    }%
    \gplgaddtomacro\gplfronttext{%
      \csname LTb\endcsname%
      \put(24,1554){\rotatebox{-270}{\makebox(0,0){\strut{}$\bra{\sigma_{12}}\%$}}}%
      \put(1770,-262){\makebox(0,0){\strut{}$\Ca$}}%
      \put(1902,2711){\makebox(0,0){\strut{}}}%
    }%
    \gplbacktext
    \put(0,0){\includegraphics{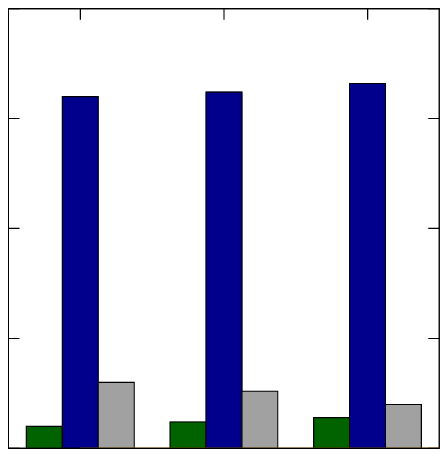}}%
    \gplfronttext
  \end{picture}%
\endgroup

%% file: tauG22.tex
\begingroup
  \makeatletter
  \providecommand\color[2][]{%
    \GenericError{(gnuplot) \space\space\space\@spaces}{%
      Package color not loaded in conjunction with
      terminal option `colourtext'%
    }{See the gnuplot documentation for explanation.%
    }{Either use 'blacktext' in gnuplot or load the package
      color.sty in LaTeX.}%
    \renewcommand\color[2][]{}%
  }%
  \providecommand\includegraphics[2][]{%
    \GenericError{(gnuplot) \space\space\space\@spaces}{%
      Package graphicx or graphics not loaded%
    }{See the gnuplot documentation for explanation.%
    }{The gnuplot epslatex terminal needs graphicx.sty or graphics.sty.}%
    \renewcommand\includegraphics[2][]{}%
  }%
  \providecommand\rotatebox[2]{#2}%
  \@ifundefined{ifGPcolor}{%
    \newif\ifGPcolor
    \GPcolortrue
  }{}%
  \@ifundefined{ifGPblacktext}{%
    \newif\ifGPblacktext
    \GPblacktexttrue
  }{}%
  \let\gplgaddtomacro\g@addto@macro
  \gdef\gplbacktext{}%
  \gdef\gplfronttext{}%
  \makeatother
  \ifGPblacktext
    \def\colorrgb#1{}%
    \def\colorgray#1{}%
  \else
    \ifGPcolor
      \def\colorrgb#1{\color[rgb]{#1}}%
      \def\colorgray#1{\color[gray]{#1}}%
      \expandafter\def\csname LTw\endcsname{\color{white}}%
      \expandafter\def\csname LTb\endcsname{\color{black}}%
      \expandafter\def\csname LTa\endcsname{\color{black}}%
      \expandafter\def\csname LT0\endcsname{\color[rgb]{1,0,0}}%
      \expandafter\def\csname LT1\endcsname{\color[rgb]{0,1,0}}%
      \expandafter\def\csname LT2\endcsname{\color[rgb]{0,0,1}}%
      \expandafter\def\csname LT3\endcsname{\color[rgb]{1,0,1}}%
      \expandafter\def\csname LT4\endcsname{\color[rgb]{0,1,1}}%
      \expandafter\def\csname LT5\endcsname{\color[rgb]{1,1,0}}%
      \expandafter\def\csname LT6\endcsname{\color[rgb]{0,0,0}}%
      \expandafter\def\csname LT7\endcsname{\color[rgb]{1,0.3,0}}%
      \expandafter\def\csname LT8\endcsname{\color[rgb]{0.5,0.5,0.5}}%
    \else
      \def\colorrgb#1{\color{black}}%
      \def\colorgray#1{\color[gray]{#1}}%
      \expandafter\def\csname LTw\endcsname{\color{white}}%
      \expandafter\def\csname LTb\endcsname{\color{black}}%
      \expandafter\def\csname LTa\endcsname{\color{black}}%
      \expandafter\def\csname LT0\endcsname{\color{black}}%
      \expandafter\def\csname LT1\endcsname{\color{black}}%
      \expandafter\def\csname LT2\endcsname{\color{black}}%
      \expandafter\def\csname LT3\endcsname{\color{black}}%
      \expandafter\def\csname LT4\endcsname{\color{black}}%
      \expandafter\def\csname LT5\endcsname{\color{black}}%
      \expandafter\def\csname LT6\endcsname{\color{black}}%
      \expandafter\def\csname LT7\endcsname{\color{black}}%
      \expandafter\def\csname LT8\endcsname{\color{black}}%
    \fi
  \fi
    \setlength{\unitlength}{0.0500bp}%
    \ifx\gptboxheight\undefined%
      \newlength{\gptboxheight}%
      \newlength{\gptboxwidth}%
      \newsavebox{\gptboxtext}%
    \fi%
    \setlength{\fboxrule}{0.5pt}%
    \setlength{\fboxsep}{1pt}%
\begin{picture}(3310.00,2880.00)%
\definecolor{gpBackground}{rgb}{1.000, 1.000, 1.000}%
\put(0,0){\colorbox{gpBackground}{\makebox(3310.00,2880.00)[]{}}}%
    \gplgaddtomacro\gplbacktext{%
      \csname LTb\endcsname%
      \put(530,288){\makebox(0,0)[r]{\strut{}$0$}}%
      \put(530,1555){\makebox(0,0)[r]{\strut{}$0.5$}}%
      \put(530,2821){\makebox(0,0)[r]{\strut{}$1$}}%
      \put(662,68){\makebox(0,0){\strut{}$0$}}%
      \put(1903,68){\makebox(0,0){\strut{}$0.5$}}%
      \put(3143,68){\makebox(0,0){\strut{}$1$}}%
    }%
    \gplgaddtomacro\gplfronttext{%
      \csname LTb\endcsname%
      \put(24,1554){\rotatebox{-270}{\makebox(0,0){\strut{}$\bra{\sigma_{12}}$}}}%
      \put(1770,-262){\makebox(0,0){\strut{}$y/h$}}%
      \put(1902,2711){\makebox(0,0){\strut{}}}%
    }%
    \gplbacktext
    \put(0,0){\includegraphics{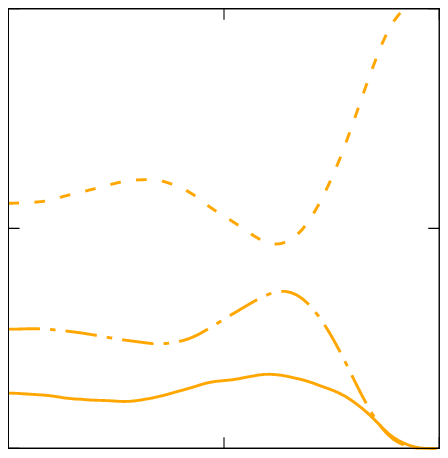}}%
    \gplfronttext
  \end{picture}%
\endgroup

%% file: tauhist2.tex
\begingroup
  \makeatletter
  \providecommand\color[2][]{%
    \GenericError{(gnuplot) \space\space\space\@spaces}{%
      Package color not loaded in conjunction with
      terminal option `colourtext'%
    }{See the gnuplot documentation for explanation.%
    }{Either use 'blacktext' in gnuplot or load the package
      color.sty in LaTeX.}%
    \renewcommand\color[2][]{}%
  }%
  \providecommand\includegraphics[2][]{%
    \GenericError{(gnuplot) \space\space\space\@spaces}{%
      Package graphicx or graphics not loaded%
    }{See the gnuplot documentation for explanation.%
    }{The gnuplot epslatex terminal needs graphicx.sty or graphics.sty.}%
    \renewcommand\includegraphics[2][]{}%
  }%
  \providecommand\rotatebox[2]{#2}%
  \@ifundefined{ifGPcolor}{%
    \newif\ifGPcolor
    \GPcolortrue
  }{}%
  \@ifundefined{ifGPblacktext}{%
    \newif\ifGPblacktext
    \GPblacktexttrue
  }{}%
  \let\gplgaddtomacro\g@addto@macro
  \gdef\gplbacktext{}%
  \gdef\gplfronttext{}%
  \makeatother
  \ifGPblacktext
    \def\colorrgb#1{}%
    \def\colorgray#1{}%
  \else
    \ifGPcolor
      \def\colorrgb#1{\color[rgb]{#1}}%
      \def\colorgray#1{\color[gray]{#1}}%
      \expandafter\def\csname LTw\endcsname{\color{white}}%
      \expandafter\def\csname LTb\endcsname{\color{black}}%
      \expandafter\def\csname LTa\endcsname{\color{black}}%
      \expandafter\def\csname LT0\endcsname{\color[rgb]{1,0,0}}%
      \expandafter\def\csname LT1\endcsname{\color[rgb]{0,1,0}}%
      \expandafter\def\csname LT2\endcsname{\color[rgb]{0,0,1}}%
      \expandafter\def\csname LT3\endcsname{\color[rgb]{1,0,1}}%
      \expandafter\def\csname LT4\endcsname{\color[rgb]{0,1,1}}%
      \expandafter\def\csname LT5\endcsname{\color[rgb]{1,1,0}}%
      \expandafter\def\csname LT6\endcsname{\color[rgb]{0,0,0}}%
      \expandafter\def\csname LT7\endcsname{\color[rgb]{1,0.3,0}}%
      \expandafter\def\csname LT8\endcsname{\color[rgb]{0.5,0.5,0.5}}%
    \else
      \def\colorrgb#1{\color{black}}%
      \def\colorgray#1{\color[gray]{#1}}%
      \expandafter\def\csname LTw\endcsname{\color{white}}%
      \expandafter\def\csname LTb\endcsname{\color{black}}%
      \expandafter\def\csname LTa\endcsname{\color{black}}%
      \expandafter\def\csname LT0\endcsname{\color{black}}%
      \expandafter\def\csname LT1\endcsname{\color{black}}%
      \expandafter\def\csname LT2\endcsname{\color{black}}%
      \expandafter\def\csname LT3\endcsname{\color{black}}%
      \expandafter\def\csname LT4\endcsname{\color{black}}%
      \expandafter\def\csname LT5\endcsname{\color{black}}%
      \expandafter\def\csname LT6\endcsname{\color{black}}%
      \expandafter\def\csname LT7\endcsname{\color{black}}%
      \expandafter\def\csname LT8\endcsname{\color{black}}%
    \fi
  \fi
    \setlength{\unitlength}{0.0500bp}%
    \ifx\gptboxheight\undefined%
      \newlength{\gptboxheight}%
      \newlength{\gptboxwidth}%
      \newsavebox{\gptboxtext}%
    \fi%
    \setlength{\fboxrule}{0.5pt}%
    \setlength{\fboxsep}{1pt}%
\begin{picture}(3310.00,2880.00)%
\definecolor{gpBackground}{rgb}{1.000, 1.000, 1.000}%
\put(0,0){\colorbox{gpBackground}{\makebox(3310.00,2880.00)[]{}}}%
    \gplgaddtomacro\gplbacktext{%
      \csname LTb\endcsname%
      \put(530,288){\makebox(0,0)[r]{\strut{}$0$}}%
      \put(530,921){\makebox(0,0)[r]{\strut{}$25$}}%
      \put(530,1555){\makebox(0,0)[r]{\strut{}$50$}}%
      \put(530,2188){\makebox(0,0)[r]{\strut{}$75$}}%
      \put(530,2821){\makebox(0,0)[r]{\strut{}$100$}}%
      \put(1076,68){\makebox(0,0){\strut{}0.1}}%
      \put(1903,68){\makebox(0,0){\strut{}0.2}}%
      \put(2730,68){\makebox(0,0){\strut{}0.4}}%
    }%
    \gplgaddtomacro\gplfronttext{%
      \csname LTb\endcsname%
      \put(24,1554){\rotatebox{-270}{\makebox(0,0){\strut{}$\bra{\sigma_{12}}\%$}}}%
      \put(1770,-262){\makebox(0,0){\strut{}$\Ca$}}%
      \put(1902,2711){\makebox(0,0){\strut{}}}%
    }%
    \gplbacktext
    \put(0,0){\includegraphics{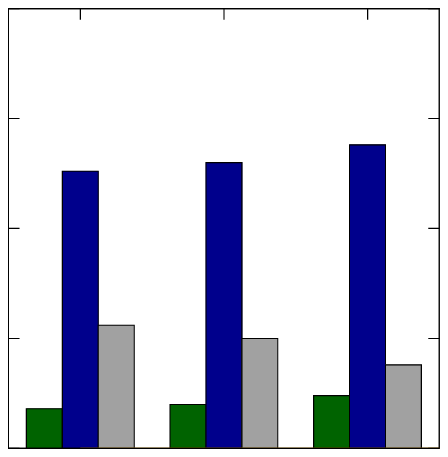}}%
    \gplfronttext
  \end{picture}%
\endgroup

%% file: tauG33.tex
\begingroup
  \makeatletter
  \providecommand\color[2][]{%
    \GenericError{(gnuplot) \space\space\space\@spaces}{%
      Package color not loaded in conjunction with
      terminal option `colourtext'%
    }{See the gnuplot documentation for explanation.%
    }{Either use 'blacktext' in gnuplot or load the package
      color.sty in LaTeX.}%
    \renewcommand\color[2][]{}%
  }%
  \providecommand\includegraphics[2][]{%
    \GenericError{(gnuplot) \space\space\space\@spaces}{%
      Package graphicx or graphics not loaded%
    }{See the gnuplot documentation for explanation.%
    }{The gnuplot epslatex terminal needs graphicx.sty or graphics.sty.}%
    \renewcommand\includegraphics[2][]{}%
  }%
  \providecommand\rotatebox[2]{#2}%
  \@ifundefined{ifGPcolor}{%
    \newif\ifGPcolor
    \GPcolortrue
  }{}%
  \@ifundefined{ifGPblacktext}{%
    \newif\ifGPblacktext
    \GPblacktexttrue
  }{}%
  \let\gplgaddtomacro\g@addto@macro
  \gdef\gplbacktext{}%
  \gdef\gplfronttext{}%
  \makeatother
  \ifGPblacktext
    \def\colorrgb#1{}%
    \def\colorgray#1{}%
  \else
    \ifGPcolor
      \def\colorrgb#1{\color[rgb]{#1}}%
      \def\colorgray#1{\color[gray]{#1}}%
      \expandafter\def\csname LTw\endcsname{\color{white}}%
      \expandafter\def\csname LTb\endcsname{\color{black}}%
      \expandafter\def\csname LTa\endcsname{\color{black}}%
      \expandafter\def\csname LT0\endcsname{\color[rgb]{1,0,0}}%
      \expandafter\def\csname LT1\endcsname{\color[rgb]{0,1,0}}%
      \expandafter\def\csname LT2\endcsname{\color[rgb]{0,0,1}}%
      \expandafter\def\csname LT3\endcsname{\color[rgb]{1,0,1}}%
      \expandafter\def\csname LT4\endcsname{\color[rgb]{0,1,1}}%
      \expandafter\def\csname LT5\endcsname{\color[rgb]{1,1,0}}%
      \expandafter\def\csname LT6\endcsname{\color[rgb]{0,0,0}}%
      \expandafter\def\csname LT7\endcsname{\color[rgb]{1,0.3,0}}%
      \expandafter\def\csname LT8\endcsname{\color[rgb]{0.5,0.5,0.5}}%
    \else
      \def\colorrgb#1{\color{black}}%
      \def\colorgray#1{\color[gray]{#1}}%
      \expandafter\def\csname LTw\endcsname{\color{white}}%
      \expandafter\def\csname LTb\endcsname{\color{black}}%
      \expandafter\def\csname LTa\endcsname{\color{black}}%
      \expandafter\def\csname LT0\endcsname{\color{black}}%
      \expandafter\def\csname LT1\endcsname{\color{black}}%
      \expandafter\def\csname LT2\endcsname{\color{black}}%
      \expandafter\def\csname LT3\endcsname{\color{black}}%
      \expandafter\def\csname LT4\endcsname{\color{black}}%
      \expandafter\def\csname LT5\endcsname{\color{black}}%
      \expandafter\def\csname LT6\endcsname{\color{black}}%
      \expandafter\def\csname LT7\endcsname{\color{black}}%
      \expandafter\def\csname LT8\endcsname{\color{black}}%
    \fi
  \fi
    \setlength{\unitlength}{0.0500bp}%
    \ifx\gptboxheight\undefined%
      \newlength{\gptboxheight}%
      \newlength{\gptboxwidth}%
      \newsavebox{\gptboxtext}%
    \fi%
    \setlength{\fboxrule}{0.5pt}%
    \setlength{\fboxsep}{1pt}%
\begin{picture}(3310.00,2880.00)%
\definecolor{gpBackground}{rgb}{1.000, 1.000, 1.000}%
\put(0,0){\colorbox{gpBackground}{\makebox(3310.00,2880.00)[]{}}}%
    \gplgaddtomacro\gplbacktext{%
      \csname LTb\endcsname%
      \put(530,288){\makebox(0,0)[r]{\strut{}$0$}}%
      \put(530,1555){\makebox(0,0)[r]{\strut{}$0.5$}}%
      \put(530,2821){\makebox(0,0)[r]{\strut{}$1$}}%
      \put(662,68){\makebox(0,0){\strut{}$0$}}%
      \put(1903,68){\makebox(0,0){\strut{}$0.5$}}%
      \put(3143,68){\makebox(0,0){\strut{}$1$}}%
    }%
    \gplgaddtomacro\gplfronttext{%
      \csname LTb\endcsname%
      \put(24,1554){\rotatebox{-270}{\makebox(0,0){\strut{}$\bra{\sigma_{12}}$}}}%
      \put(1770,-262){\makebox(0,0){\strut{}$y/h$}}%
      \put(1902,2711){\makebox(0,0){\strut{}}}%
    }%
    \gplbacktext
    \put(0,0){\includegraphics{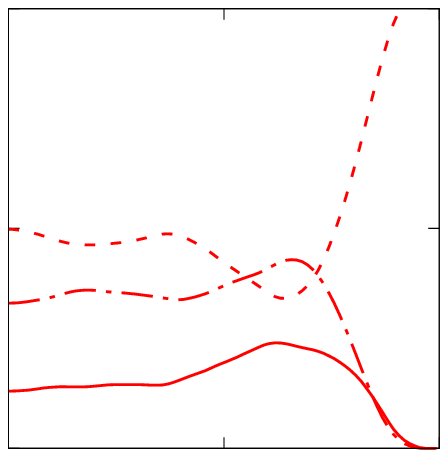}}%
    \gplfronttext
  \end{picture}%
\endgroup

%% file: tauhist3.tex
\begingroup
  \makeatletter
  \providecommand\color[2][]{%
    \GenericError{(gnuplot) \space\space\space\@spaces}{%
      Package color not loaded in conjunction with
      terminal option `colourtext'%
    }{See the gnuplot documentation for explanation.%
    }{Either use 'blacktext' in gnuplot or load the package
      color.sty in LaTeX.}%
    \renewcommand\color[2][]{}%
  }%
  \providecommand\includegraphics[2][]{%
    \GenericError{(gnuplot) \space\space\space\@spaces}{%
      Package graphicx or graphics not loaded%
    }{See the gnuplot documentation for explanation.%
    }{The gnuplot epslatex terminal needs graphicx.sty or graphics.sty.}%
    \renewcommand\includegraphics[2][]{}%
  }%
  \providecommand\rotatebox[2]{#2}%
  \@ifundefined{ifGPcolor}{%
    \newif\ifGPcolor
    \GPcolortrue
  }{}%
  \@ifundefined{ifGPblacktext}{%
    \newif\ifGPblacktext
    \GPblacktexttrue
  }{}%
  \let\gplgaddtomacro\g@addto@macro
  \gdef\gplbacktext{}%
  \gdef\gplfronttext{}%
  \makeatother
  \ifGPblacktext
    \def\colorrgb#1{}%
    \def\colorgray#1{}%
  \else
    \ifGPcolor
      \def\colorrgb#1{\color[rgb]{#1}}%
      \def\colorgray#1{\color[gray]{#1}}%
      \expandafter\def\csname LTw\endcsname{\color{white}}%
      \expandafter\def\csname LTb\endcsname{\color{black}}%
      \expandafter\def\csname LTa\endcsname{\color{black}}%
      \expandafter\def\csname LT0\endcsname{\color[rgb]{1,0,0}}%
      \expandafter\def\csname LT1\endcsname{\color[rgb]{0,1,0}}%
      \expandafter\def\csname LT2\endcsname{\color[rgb]{0,0,1}}%
      \expandafter\def\csname LT3\endcsname{\color[rgb]{1,0,1}}%
      \expandafter\def\csname LT4\endcsname{\color[rgb]{0,1,1}}%
      \expandafter\def\csname LT5\endcsname{\color[rgb]{1,1,0}}%
      \expandafter\def\csname LT6\endcsname{\color[rgb]{0,0,0}}%
      \expandafter\def\csname LT7\endcsname{\color[rgb]{1,0.3,0}}%
      \expandafter\def\csname LT8\endcsname{\color[rgb]{0.5,0.5,0.5}}%
    \else
      \def\colorrgb#1{\color{black}}%
      \def\colorgray#1{\color[gray]{#1}}%
      \expandafter\def\csname LTw\endcsname{\color{white}}%
      \expandafter\def\csname LTb\endcsname{\color{black}}%
      \expandafter\def\csname LTa\endcsname{\color{black}}%
      \expandafter\def\csname LT0\endcsname{\color{black}}%
      \expandafter\def\csname LT1\endcsname{\color{black}}%
      \expandafter\def\csname LT2\endcsname{\color{black}}%
      \expandafter\def\csname LT3\endcsname{\color{black}}%
      \expandafter\def\csname LT4\endcsname{\color{black}}%
      \expandafter\def\csname LT5\endcsname{\color{black}}%
      \expandafter\def\csname LT6\endcsname{\color{black}}%
      \expandafter\def\csname LT7\endcsname{\color{black}}%
      \expandafter\def\csname LT8\endcsname{\color{black}}%
    \fi
  \fi
    \setlength{\unitlength}{0.0500bp}%
    \ifx\gptboxheight\undefined%
      \newlength{\gptboxheight}%
      \newlength{\gptboxwidth}%
      \newsavebox{\gptboxtext}%
    \fi%
    \setlength{\fboxrule}{0.5pt}%
    \setlength{\fboxsep}{1pt}%
\begin{picture}(3310.00,2880.00)%
\definecolor{gpBackground}{rgb}{1.000, 1.000, 1.000}%
\put(0,0){\colorbox{gpBackground}{\makebox(3310.00,2880.00)[]{}}}%
    \gplgaddtomacro\gplbacktext{%
      \csname LTb\endcsname%
      \put(530,288){\makebox(0,0)[r]{\strut{}$0$}}%
      \put(530,921){\makebox(0,0)[r]{\strut{}$25$}}%
      \put(530,1555){\makebox(0,0)[r]{\strut{}$50$}}%
      \put(530,2188){\makebox(0,0)[r]{\strut{}$75$}}%
      \put(530,2821){\makebox(0,0)[r]{\strut{}$100$}}%
      \put(1076,68){\makebox(0,0){\strut{}0.1}}%
      \put(1903,68){\makebox(0,0){\strut{}0.2}}%
      \put(2730,68){\makebox(0,0){\strut{}0.4}}%
    }%
    \gplgaddtomacro\gplfronttext{%
      \csname LTb\endcsname%
      \put(24,1554){\rotatebox{-270}{\makebox(0,0){\strut{}$\bra{\sigma_{12}}\%$}}}%
      \put(1770,-262){\makebox(0,0){\strut{}$\Ca$}}%
      \put(1902,2711){\makebox(0,0){\strut{}}}%
    }%
    \gplbacktext
    \put(0,0){\includegraphics{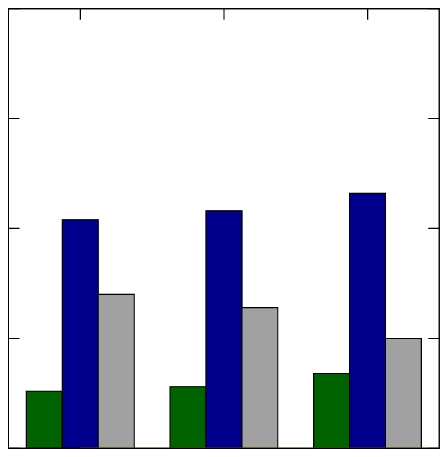}}%
    \gplfronttext
  \end{picture}%
\endgroup